\documentclass[12pt]{article}
\usepackage{amsmath,amsfonts,graphicx,yfonts,mathabx}
\usepackage[nosort]{cite}

\usepackage{amssymb}
\usepackage{epsfig, float}

\def\baselinestretch{1.1}
\usepackage[T1]{fontenc} 
\usepackage[utf8]{inputenc}

\usepackage{tikz}
\usetikzlibrary{decorations.pathmorphing}
\usepackage{capt-of}
\usepackage{caption}

\newcommand{\be}{\begin{equation}}
\newcommand{\ee}{\end{equation}}
\newcommand{\bea}{\begin{eqnarray}}
\newcommand{\eea}{\end{eqnarray}}

\newcommand{\mt}[1]{\textrm{\tiny #1}}

\def\rh {r_\mt{H}}

\renewcommand{\title}[1]{\vbox{\center\LARGE{#1}}\vspace{3mm}}
\renewcommand{\author}[1]{\vbox{\center#1}\vspace{3mm}}
\newcommand{\address}[1]{\vbox{\center\em#1}}
\newcommand{\email}[1]{\vbox{\center\tt#1}\vspace{3mm}}

\begin{document}
\begin{titlepage}
\begin{center}
\rightline{\tt}
\vskip 2.5cm
\title{Recent developments in the holographic description of quantum chaos.}
\vskip .6cm
\author{Viktor Jahnke$^{a,b}$}
\vskip -.5cm 
\address{$^{a}$Departamento de F\'isica de Altas Energias, Instituto de Ciencias Nucleares, Universidad Nacional Aut\' onoma de M\' exico\\ Apartado Postal 70-543, CDMX 04510, M\'exico}
\vskip -.1cm

\address{$^{b}$School of Physics and Chemistry, Gwangju Institute of Science and Technology, Gwangju 61005, Korea}
\vskip -.1cm

\email{viktor.jahnke@correo.nucleares.unam.mx\\ viktorjahnke@gist.ac.kr}

\end{center}
\vskip 3cm

\abstract{ \small
We review recent developments encompassing the description of quantum chaos in holography. We discuss the characterization of quantum chaos based on the late time vanishing of out-of-time-order correlators and explain how this is realized in the dual gravitational description. We also review the connections of chaos with the spreading of quantum entanglement and diffusion phenomena.

}

\vfill
\end{titlepage}
%%\keywords{Gauge-gravity correspondence, Holography and quark-gluon plasmas}  

%%\emailAdd{viktor.jahnke@usp.br}
%%\emailAdd{anderson.misobuchi} 

%%%%% DOCUMENT
%%\begin{document}

%%\maketitle
%%\setlength{\parskip}{3pt}
\tableofcontents

\section{Introduction} \label{sec-1}

The characterization of quantum chaos is fairly complicated. Possible approaches range from semi-classical methods to random matrix theory: in the first case one studies the semi-classical limit of a system whose classical dynamics is chaotic; in the latter approach the characterization of quantum chaos is made by comparing the spectrum of energies of the system in question to the spectrum of random matrices \cite{ullmo-2014}. Despite the insights provided by the above-mentioned approaches, a complete and more satisfactory understanding of quantum chaos remains elusive.

Surprisingly, new insights into quantum chaos have come from black holes physics! In the context of so-called gauge gravity duality \cite{duality1, duality2, duality3}, black holes in asymptotically AdS spaces are dual to strongly coupled many-body quantum systems. It was recently shown that the chaotic nature of many-body quantum systems can be diagnosed with certain out-of-time-order correlation (OTOC) functions which, in the gravitational description, are related to the collision of shock waves close to the black hole horizon \cite{BHchaos1,BHchaos2,BHchaos3,BHchaos4,Kitaev-2014}. In addition to being useful for diagnosing chaos in holographic systems and providing a deeper understanding for the inner-working mechanisms of gauge-gravity duality, OTOCs have also proved useful in characterizing chaos in more general non-holographic systems, including some simple models like the kicked-rotor \cite{Rozenbaum-2017}, the stadium billiard \cite{Rozenbaum-2018}, and the Dicke model \cite{Chavez-Carlos:2018}.

In this paper, we review the recent developments in the holographic description of quantum chaos. We discuss the characterization of quantum chaos based on the late time vanishing of OTOCs and explain how this is realized in the dual gravitational description. We also review the connections of chaos with the spreading of quantum entanglement and diffusion phenomena. We focus on the case of $d-$dimensional gravitational systems with $d \geq 3$, which excludes the case of gravity in $AdS_2$ and SYK-like models \cite{Polchinski:2016xgd,Maldacena:2016hyu, Maldacena:2016upp, Kitaev:2017awl}\footnote{Another interesting perspective on the characterization of chaos in the context of (regularized) $AdS_2/CFT_1$ is provided by \cite{Axenides:2013iwa, Axenides:2015aha, Axenides:2016nmf}.}. Also, due the lack of the author expertise, we did not cover the recent developments in the direct field theory calculations of OTOCs. This includes calculations for CFTs \cite{Roberts:2014ifa}, weakly coupled systems \cite{Stanford:2015owe,Plamadeala:2018vsr}, random unitary models \cite{Nahum:2017yvy,Khemani:2017nda,Rakovszky:2017qit} and spin chains \cite{Luitz:2017jrn,Bohrdt:2016vhv,Heyl-2018,Lin-2018,Xu:2018xfz}.

\section{A bird eye's view on classical chaos} \label{sec-2}

\begin{figure}[h!]
\centering
%\captionsetup{justification=left}

\begin{tikzpicture}[scale=1.5]
\draw [thick]  (0,0) -- (0,3);
\draw [thick]  (3,0) -- (3,3);
\draw [thick]  (0,0) -- (3,0); 
\draw [thick]  (0,3) -- (3,3);

\draw  [thick,dashed] (.5,.53) .. controls (.7,.7) and (1,1) .. (1.5,1.97); 
\draw  [thick,dashed] (.8,.53) .. controls (1,.7) and (1.3,.9) .. (2.5,1.97);

\draw [<->] (.53,.5) -- (.77,.5);
\draw [<->] (1.53,2) -- (2.47,2);

\end{tikzpicture}
\put(-140,115){$p$}
\put(-10,-8){$q$}
\put(-110,8){\small $\delta q(0)$}
\put(-50,90){\small $\delta q(t)$}

\caption{ \small Variation of a trajectory in the phase space under small modifications of the initial condition. For a chaotic system the distance between two initially nearby trajectories increases exponentially with time, i.e., $|\delta q(t)| = |\delta q(0)| e^{\lambda t}$.}
\label{fig-phaseSpace}
\end{figure}
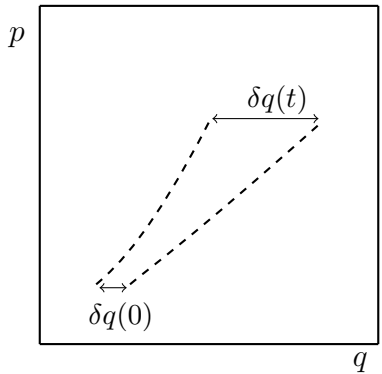

In this section we briefly review some basic aspects of classical chaos. For definiteness we consider the case of a classical thermal system with phase space denoted as ${\bf X=(q,p)}$, where ${\bf q}$ and ${\bf p}$ are multi-dimensional vectors denoting the coordinates and momenta of the phase space. We can quantify whether the system is chaotic or not by measuring the stability of a trajectory in phase space under small changes of the initial condition. Let us consider a reference trajectory in phase space, ${\bf X}(t)$, with some initial condition ${\bf X}(0)={\bf X}_0$. A small change in the initial condition ${\bf X}_0 \rightarrow {\bf X}_0 +\delta {\bf X}_0$ leads to a new trajectory $ {\bf X}(t) \rightarrow {\bf X}(t)+\delta {\bf X}(t)$. This is illustrated in figure \ref{fig-phaseSpace}. For a chaotic system, the distance between the new trajectory and the reference one increases exponentially with time
\be
|\delta {\bf X}(t)| \sim |\delta {\bf X}_0|e^{\lambda t}\,\,\, \text{or}\,\,\,\frac{\partial {\bf X}(t)}{\partial {\bf X}_0} \sim e^{\lambda t}\,,
\label{eq-deltaX}
\ee
where $\lambda$ is the so-called Lyapunov exponent. This should be contrasted with the behaviour of non-chaotic systems, in which $\delta {\bf X}(t)$ remains bounded or increase algebraically \cite{livro-chaos}. 

The exponential increase depends on the orientation of $\delta {\bf X}_0$ and this leads to a spectrum of Lyapunov exponents, $\{ \lambda_1, \lambda_2, ..., \lambda_K \}$, where $K$ is the dimensionality of the phase space.
A useful parameter characterizing the trajectory instability is
\be
\lambda_\mt{max} = \lim_{t \rightarrow \infty} \lim_{\delta {\bf X}_0 \rightarrow 0} \frac{1}{t} \log \left( \frac{\delta {\bf X}(t)}{\delta {\bf X}_0} \right)\,,
\ee
which is called the maximum Lyapunov exponent. When the above limits exist and $\lambda_\mt{max}>0$ the trajectory shows sensitive to initial conditions and the system is said to be chaotic \cite{livro-chaos}.

The chaotic behavior can be either a consequence of a complicated Hamiltonian or simply due to the contact with a thermal heat bath. This is because chaos is a common property of thermal systems. To later to make contact with black holes physics we consider the case of a classical thermal system with inverse temperature $\beta$. If $F({\bf X})$ is some function of the phase space coordinates we define its classical expectation value as
\be
\langle F \rangle_{\beta} = \frac{\int d{\bf X} e^{-\beta H({\bf X})} F({\bf X})}{\int d{\bf X} e^{-\beta H({\bf X})}}
\ee
where $H({\bf X})$ is the system's Hamiltonian.

Classical thermal systems have two exponential behaviors that  have analogues in terms of black holes physics: the Lyapunov behavior, characterizing the sensitive dependence on initial conditions; and the Ruelle behavior, characterizing the approach to thermal equilibrium \cite{Polchinski:2015cea,Garcia-Mata-2018}.

To quantify the sensitivity to initial conditions in a thermal system we need to consider thermal expectation values.
Note that (\ref{eq-deltaX}) can have either signs. To avoid cancellations in a thermal expectation values we consider the square of this derivative
\be
F(t) =\left\langle \left( \frac{\partial {\bf X}(t)}{\partial {\bf X}(0)} \right)^2 \right\rangle_{\beta}\,.
\ee
The expected behavior of this quantity is the following \cite{Douglas-PITP}
\be
F(t) \sim \sum_k c_k \, e^{2 \lambda_k t}\,,
\ee
where $c_k$ are constants and $\lambda_k$ are the Lyapunov exponents. At later times the behavior is controlled by the maximum Lyapunov exponent $F \sim e^{2\lambda_\mt{max}t}$.

The approach to thermal equilibrium or, in other words, how fast the system forgets its initial condition, can be quantified by two-point functions of the form
\be
G(t)= \langle {\bf X}(t) \,{\bf X}(0) \rangle_{\beta}-\langle {\bf X} \rangle_{\beta}^2 \,,
\ee
whose expected behavior is \cite{Douglas-PITP}
\be
G(t) \sim \sum_{j} b_j \, e^{-\mu_j t}\,,
\ee
where $b_j$ are constants and $\mu_j$ are complex parameters called Ruelle resonances. The late time behavior is controlled by the smallest Ruelle resonance $G \sim e^{-\mu_\mt{min} t}$.

\section{Some aspects of quantum chaos} \label{sec-3}
In this section, we review some aspects of quantum chaos. For a long time, the characterization of quantum chaos was made by comparing the spectrum of energies of the system in question to the spectrum of random matrices or using semiclassical methods \cite{ullmo-2014}. Here we follow a different approach, which was first proposed by Larkin and Ovchinnikov \cite{larkin} in the context of semi-classical systems, and it was recently developed by Shenker and Stanford \cite{BHchaos2,BHchaos3,BHchaos4} and by Kitaev \cite{Kitaev-2014}.

For simplicity, let us consider the case of a one-dimensional system, with phase space variables $(q,p)$. Classically, we know that $\partial q(t)/ \partial q(0)$ grows exponentially with time for a chaotic system. The quantum version of this quantity can be obtained by noting that
\be
\frac{\partial q(t)}{\partial q(0)} = \{ q(t), p(0)\}_\mt{P.B.}\,,
\ee
where $\{ q(t),p(0) \}_\mt{P.B.}$ denotes the Poisson bracket between the coordinate $q(t)$ and the momentum $p(0)$. The quantum version of $\partial q(t)/ \partial q(0)$ can then be obtained by promoting the Poisson bracket to a commutator
\be
 \{ q(t), p(0)\}_\mt{P.B.} \rightarrow \frac{1}{i \hbar} [\hat{q}(t),\hat{p}(0)]
\ee
where now $\hat{q}(t)$ and $\hat{p}(0)$ are Heisenberg operators.

We will be interested in thermal systems, so we would like to calculate the expectation value of $[\hat{q}(t),\hat{p}(0)]$ in a thermal state. However, this commutator might have either sign in a thermal expectation value and this might lead to cancellations. To overcome this problem, we consider the expectation value of the square of this commutator
\be
C(t)=\left\langle-[\hat{q}(t),\hat{p}(0)]^2 \right\rangle_{\beta}\,,
\ee
where $\beta$ is the system's inverse temperature and the overall sign is introduced to make $C(t)$ positive. More generally, one might replace $\hat{q}(t)$ and $\hat{p}(0)$ by two generic hermitian operators $V$ and $W$ and quantify chaos with the {\it double commutator}
\be
C(t) = \left\langle-[W(t),V(0)]^2 \right\rangle_{\beta}\,.
\label{eq-C(t)}
\ee
This quantity measures how much an early perturbation $V$ affects the later measurement of $W$. As chaos means sensitive dependence on initial conditions, we expect $C(t)$ to be `small' in non-chaotic system, and `large' if the dynamics is chaotic. In the following, we give a precise meaning for the adjectives `small' and `large'.

For some class of systems, the quantum behavior of $C(t)$ has a lot of similarities with the classical behavior of $\langle (\partial q(t)/\partial q(0)) \rangle_{\beta}$. However, the analogy between the classical and quantum quantities is not perfect because there is not always a good notion of a small perturbation in the quantum case (remember that classical chaos is characterized by the fact that a small perturbation in the past has important consequences in the future). If we start with some reference state and then perturb it, we easily produce a state that is orthogonal to the original state, even when we change just a few quantum numbers. Because of that, it seems unnatural to quantify the perturbation as small. Fortunately, there are some quantum systems in which the notion of a small perturbation makes perfect sense. An example is provided by systems with a large number of degrees of freedom. In this case, a perturbation involving just a few degrees of freedom is naturally a small perturbation.

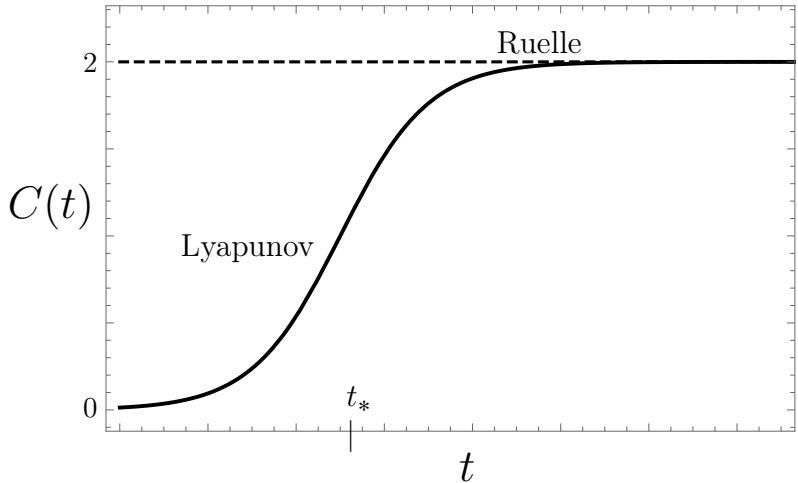
\begin{figure}[t!]
\begin{center}
\setlength{\unitlength}{1cm}
\includegraphics[width=0.6\linewidth]{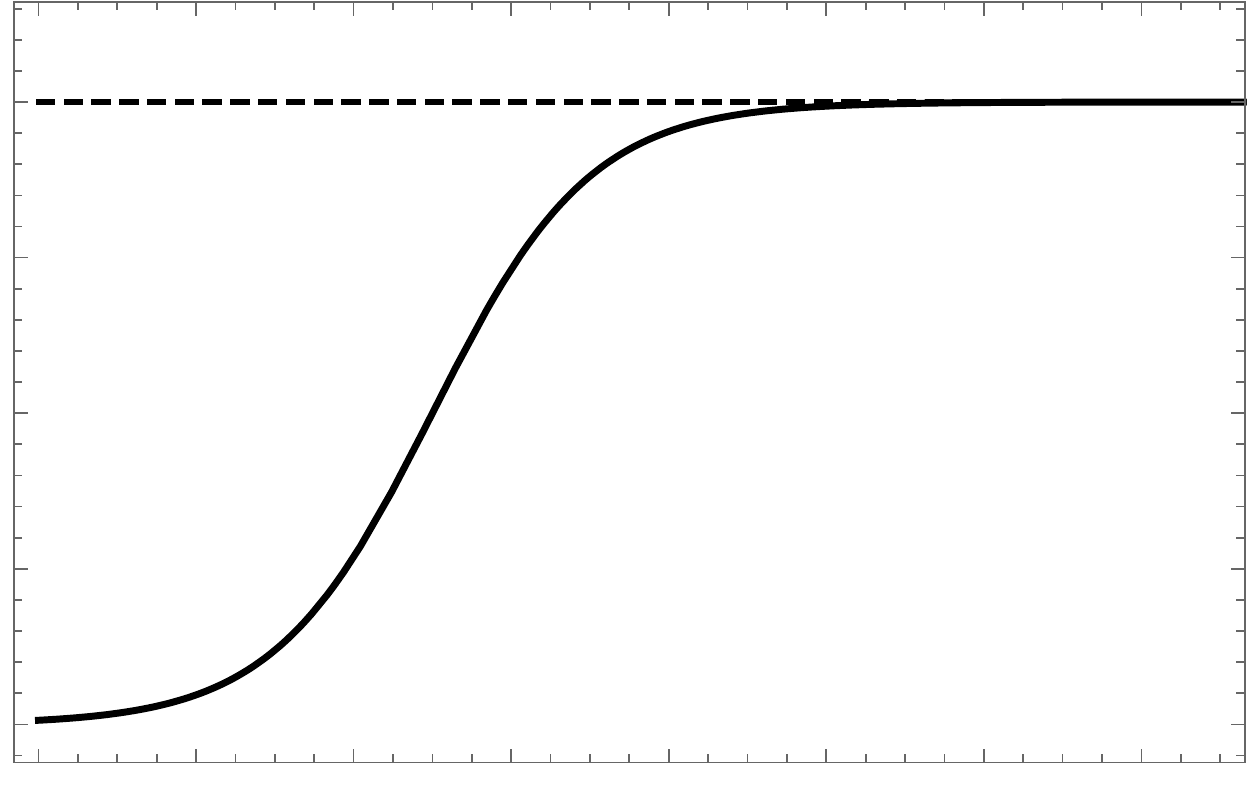}
\put(-4.5,-.5){\Large $t$}
\put(-6,+.5){$t_*$}
\put(-6,+.0){$|$}
\put(-10.5,+3.){\Large $C(t)$}
\put(-4.0,+5.2){Ruelle}
\put(-8.2,+2.5){Lyapunov}
\put(-9.5,+4.95){\small 2}
\put(-9.5,+0.35){\small 0}

\end{center}
\caption{ \small
Schematic form of $C(t)$. We indicated the regions of Lyapunov and Ruelle behavior. $C(t) \sim \mathcal{O}(1)$ at $t \sim t_*$.}
\label{fig-C(t)}
\end{figure}

For some class of chaotic systems, which include holographic systems, $C(t)$ is expected to behave as\footnote{See \cite{Xu:2018xfz,Khemani:2018sdn} for a discussion of different possible OTOC growth forms.}

\[ C(t) \sim
  \begin{cases}
    N_\mt{dof}^{-1}      & \quad \text{for }\,\, t < t_d\,,\\
    N_\mt{dof}^{-1} \exp \left( \lambda_L t\right) & \quad \text{for}\,\, t_d <\!\!< t <\!\!< t_*\,\\
    \mathcal{O}(1) & \quad \text{for}\,\, t > t_*\,,
  \end{cases}
\]
where $N_\mt{dof}$ is the number of degrees of freedom of the system. Here, we have assumed $V$ and $W$ to be unitary and hermitian operators, so that $V V= W W=1$. The exponential growth of $C(t)$ is characterized by the Lyapunov exponent\footnote{This is actually the quantum analogue of the classical Lyapunov exponent. The two quantities are not necessarily the same in the classical limit \cite{Stanford:2015owe}. Here we stick to the physicists long standing tradition of using misnomers and just refer to $\lambda_L$ as the Lyapunov expoent.} $\lambda_L$ and takes place at intermediate time scales bounded by the dissipation time $t_d$ and the scrambling time $t_*$. The dissipation time is related to the classical Ruelle resonances ($t_d \sim \mu^{-1}$) and it characterizes the exponential decay of two-point correlators, e.g., $\langle V(0) V(t) \rangle \sim e^{-t/t_d}$. The dissipation time also controls the late time behavior of $C(t)$. The scrambling time $t_* \sim \lambda_L^{-1} \log N_\mt{dof}$ is defined as the time at which $C(t)$ becomes of order $\mathcal{O}(1)$. See figure \ref{fig-C(t)}. The scrambling time controls how fast the chaotic system scrambles information. If we perturb the system with an operator that involves only a few degrees of freedom, the information about this operator will spread among the other degrees of freedom of the system. After a scrambling time, the information will be scrambled among all the degrees of freedom and the operator will have a large commutator with almost any other operator. 

To understand how the above behavior relates to chaos, we write the double commutator as
\bea
C(t)&=& \left\langle-[W(t),V(0)]^2 \right\rangle_{\beta} \\
&=& 2- 2\, \langle W(t) V(0) W(t) V(0)  \rangle_{\beta}\,,
\eea
where we made the assumption that $W$ and $V$ are hermitian and unitary operators. Note that all the relevant information about $C(t)$ is contained in the OTOC
\be
\text{OTO}(t)=\left\langle W(t) V(0) W(t) V(0) \right\rangle\,.
\ee
The fact that $C(t)$ approaches 2 at later times implies that the OTO$(t)$ should vanish in that limit. To understand why this is related to chaos we think of OTO$(t)$ as an inner-product of two states
\be 
\text{OTO}(t)=\langle \psi_2 | \psi_1 \rangle\,,
\ee
where
\be
|\psi_1 \rangle = W(-t)V(0)|\beta \rangle\,, \,\,\,\, |\psi_2 \rangle = V(0)W(-t)|\beta \rangle
\ee
where $| \beta \rangle$ is some thermal state and we replace $t \rightarrow -t$ to make easier the comparison with black holes physics.

If $[V(0),W(t)] \approx 0$ for any value of $t$, the two states are approximately the same, and $ \langle \psi_1 | \psi_2 \rangle\ \approx 1$, implying $C(t) \approx 0$. That means the system displays no chaos - the early measurement of $V$ has no effect on the later measurement of $W$. If, on the other hand, $[V(0),W(t)] \neq 0$, the states $|\psi_1 \rangle$ and $|\psi_2 \rangle $ will have a small superposition $ \langle \psi_1 | \psi_2 \rangle\ \approx 0$, implying $C(t) \approx 2$. That means that $V$ has a large effect on the later measurement of $W$.

\begin{figure}[H]
\begin{center}
\includegraphics[width=10.5cm]{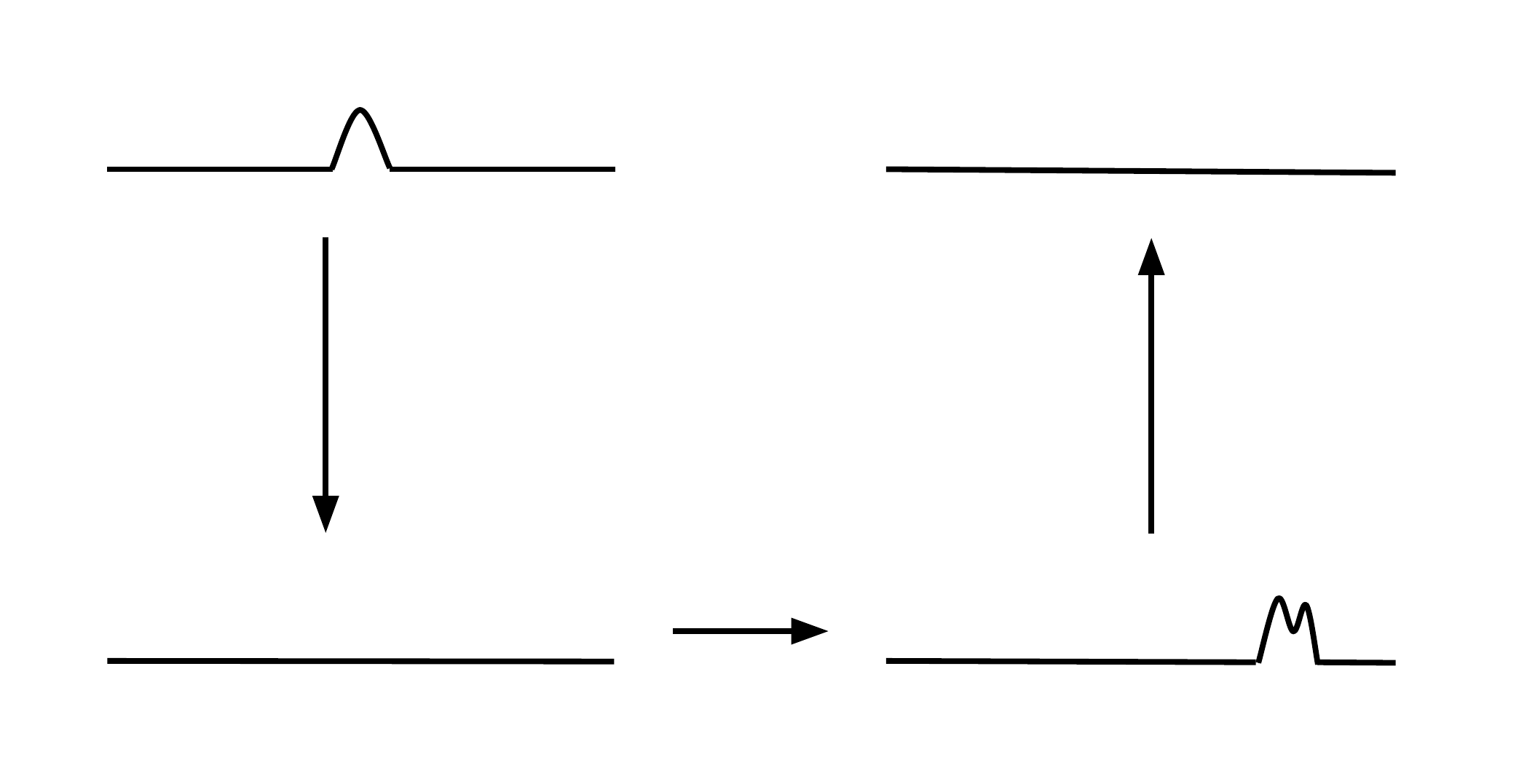}
\put(-200,105){\small $V(0)| \beta \rangle$}
\put(-220,8){\small $e^{iHt}V(0)| \beta \rangle$}
\put(-60,8){\small $W(0)e^{iHt}V(0)| \beta \rangle$}
\put(-70,105){\small $e^{-iHt}W(0)e^{iHt}V(0)| \beta \rangle$}
\put(-310,118){\small $t=0$}
\put(-320,20){\small $-t<0$}

\end{center}
\caption{ \small Construction of the state $W(-t)V(0)| \beta \rangle$. For a chaotic system the perturbation $V$ fails to re-materialize at $t=0$. In a non-chaotic system we expect the perturbation $V$ to re-materialize at $t=0$.}
\label{fig-WVstate1}
\end{figure}

In figure \ref{fig-WVstate1} we construct the states $|\psi_1 \rangle$ and $|\psi_2 \rangle$ and explain why $ \langle \psi_1 | \psi_2 \rangle\ \approx 0$  for large $t$ means chaos. Let us start by constructing the state $|\psi_1 \rangle =W(-t)V(0)|\beta \rangle = e^{-iHt}W(0)e^{iHt}V(0)|\beta \rangle$. The unperturbed thermal state is represented by a horizontal line. We initially consider the state $V(0)| \beta \rangle$, which is the thermal state perturbed by $V$. If we evolve the system backward in time (applying the operator $e^{iHt}$) for some time which is larger than the dissipation time, the system will thermalize and it will no longer display the perturbation $V$. After that, we apply the operator $W$, which should be thought of as a small perturbation, and then we evolve the system forwards in time (applying the operator $e^{-iHt}$). The final result of this set of operations depends on the nature of the system. If the system is chaotic, the perturbation $W$ will have a large effect after a scrambling time, and the perturbation that was present at $t=0$ will no longer re-materialize. This is illustrated in figure \ref{fig-WVstate1}. In contrast, for a non-chaotic system, the perturbation $W$ will have little effect on the system at later times, and the perturbation $V$ will (at least partially) re-materialize at $t=0$. 

\begin{figure}[H]
\begin{center}
\includegraphics[width=12.5cm]{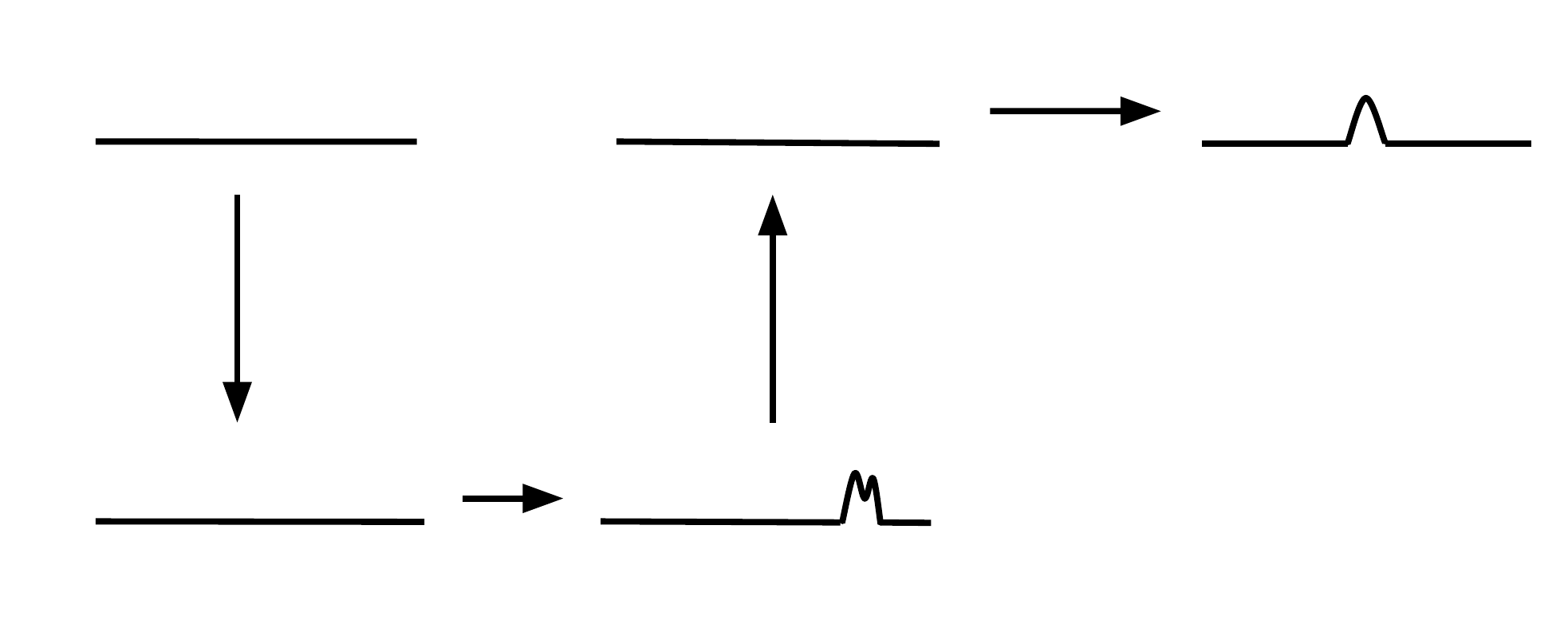}
\put(-260,95){\small $| \beta \rangle$}
\put(-280,10){\small $e^{iHt}| \beta \rangle$}
\put(-170,10){\small $W(0)e^{iHt}| \beta \rangle$}
\put(-175,95){\small $e^{-iHt}W(0)e^{iHt}| \beta \rangle$}
\put(-60,95){\small $V(0)e^{-iHt}W(0)e^{iHt}| \beta \rangle$}
\put(-370,108){\small $t=0$}
\put(-375,22){\small $-t<0$}

\end{center}
\caption{\small Construction of the state $V(0)W(-t)| \beta \rangle$. By construction, this state displays the perturbation $V$ at $t=0$.}
\label{fig-WVstate2}
\end{figure}
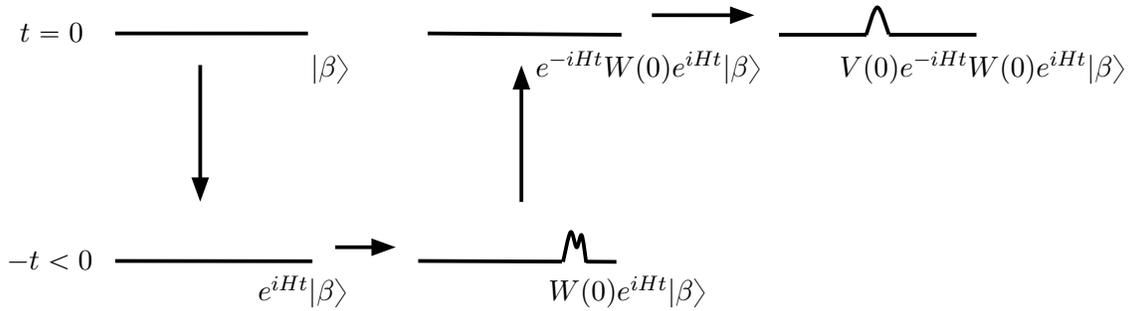

We now construct the state $|\psi_2 \rangle =V(0)W(-t)|\beta \rangle = V(0)e^{-iHt}W(0)e^{iHt}|\beta \rangle$. This is illustrated in figure \ref{fig-WVstate2}. We start with the thermal state $|\beta \rangle$ and then we evolve this state backwards in time $e^{iHt}|\beta \rangle$. After that we apply the operator $W$ and then we evolve the system forwards in time, obtaining the state $e^{-iHt}W(0)e^{iHt}|\beta \rangle$. Finally, we apply the operator $V$, obtaining the state $V(0)W(-t)|\beta \rangle$. Note that, by construction, this state displays the perturbation $V$ at $t=0$, while the state $W(-t)V(0)|\beta \rangle$ does not. As a consequence, the two states are expected to have a small superposition $\langle W(-t)V(0)|W(-t)V(0) \rangle_{\beta} \approx 0$. This should be contrasted to the case where the system is not chaotic. In this case the perturbation $V$ re-materializes at $t=0$, and the states $|\psi_1 \rangle$ and $| \psi_2 \rangle$ have a large superposition, i.e., $\langle W(-t)V(0)|W(-t)V(0) \rangle_{\beta} \approx 1$.

In this construction we assumed the operators $V(0)$ and $W(-t)$ to be separated by a scrambling time, i.e., $ |t| > t_*$. This is important because, at earlier times, the two operators, which in general involve different degrees of freedom of the system, generically commute. The operators manage to have a non-zero commutator at later times because of the phenomenon of operation growth that we will describe in the next section.

\subsection{Operator growth and scrambling}\label{sec-opgrow}

The operators $V$ and $W$ act generically at different parts of the physical system, yet they can have a non-zero commutator at later times. This is possible because in chaotic systems the time evolution of an operator makes it more and more complicated, involving and increasing number of degrees of freedom. As a result, an operator that initially involves just a few degrees of freedom becomes delocalized over a region that grows with time. The growth of the operator $W(t)$ is maybe more evident from the point of view of the Baker-Campbell-Hausdorff (BCH) formula, in terms of which we can write
\be
W(t)=e^{iHt}W(0)e^{-iHt}=\sum_{k=0}^{\infty} \frac{(-it)^k}{k!}[H[H,...[H,W(0)]...]\,]\,.
\ee
From the above formula it is clear that, at each order in $t$ there is a more complicated contribution to $W(t)$. In chaotic systems the operator becomes more and more delocalized as the time evolves, and it eventually becomes delocalized over the entire system. The time scale at which this occurs is the so-called scrambling time $t_*$. After the scrambling time the operator $W(t)$ manages to have a non-zero and large commutator with almost any other operator, even operators involving only a few degrees of freedom.

This can be clearly illustrated in the case of a spin-chain. Let us follow \cite{BHchaos3} and consider an Ising-like model with Hamiltonian
\be
H=-\sum_{i}\left(Z_i Z_{i+1}+g X_i + h Z_i \right)\,,
\ee
where $X_i, Y_i$ and $Z_i$ denote Pauli matrices acting on the $i$th site of the spin chain. The above system is integrable if  we take $g=1$ and $h=0$, but it is strongly chaotic if we choose $g=-1.05$ and $h=0.5$.

To illustrate the concept of scrambling, we consider the time evolution of the operator $Z_1$. Using the BCH formula we can write
\be
Z_1(t)=Z_1-it[H,Z_1]-\frac{t^2}{2!}[H,[H,Z_1]\,]+\frac{i t^3}{3!} [H,[H,[H,Z_1]\,]\,]+...
\label{eq-BCHseries}
\ee
Ignoring multiplicative constants and signs we can write the above terms (schematically) as

\begin{align}
 [H,Z_1]  \sim &   Y_1  \\
 [H,[ H,Z_1]\,]  \sim & Y_1+X_1 Z_2 \nonumber \\
 [H,[H,[H,Z_1]]]  \sim &  Y_1+Y_2X_1+Y_1Z_2 \nonumber \\
 [H,[H,[H,[H,Z_1]]]]  \sim & X_1+Y_1+Z_1+X_1 X_2+Y_1 Y_2+Z_1 Z_2+ X_1 Z_2+ \nonumber \\
 & + Z_3 Y_1 + Y_1 Z_2 Y_2+Z_1 X_2 X_1+X_2 Z_3 X_1 \nonumber
\end{align}
As the the time evolves higher order terms become important in the series (\ref{eq-BCHseries}), and the operator $Z_1(t)$ becomes more and more complicated, involving terms in an increasing number of sites. For large enough $t$ the operator will involve all the sites of the spin chain and it will manage to have a non-zero commutator with a Pauli operator in any other site of the system. In this situation the information about $Z_1$ is essentially scramble among all the degrees of freedom of the system. As discussed before, this occurs after a scrambling time. Above this time the double commutator $C(t)$ saturates to a constant value. This should be contrasted to what happens for an integrable system. In this case the operator grows, but it also decreases at later times. In the chaotic case, the operator remains large at later times \cite{BHchaos3}.  

\subsection{Probing chaos with local operators}
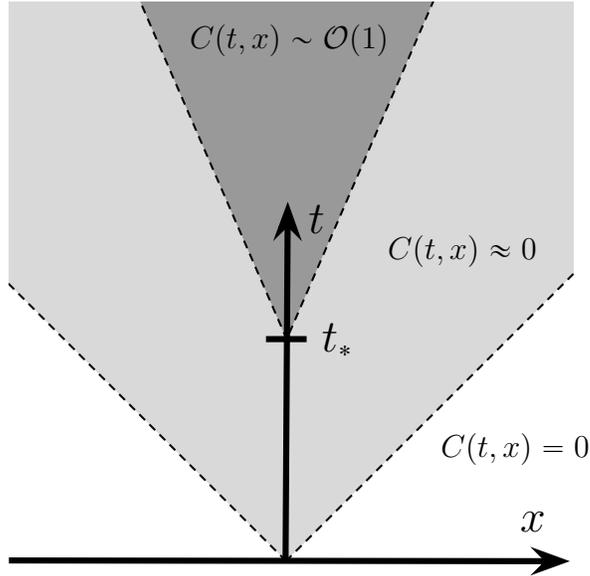
\begin{figure}[H]
\begin{center}
\includegraphics[width=7.5cm]{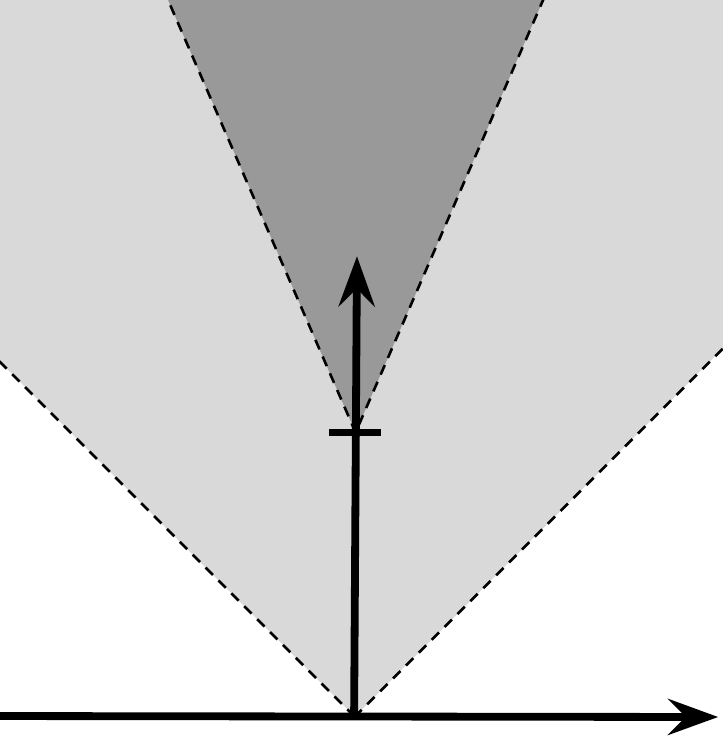}
\put(-100,130){\Large $t$}
\put(-95,85){\Large $t_*$}
\put(-20,17){\Large $x$}
\put(-50,45){$C(t,x)=0$}
\put(-70,120){$C(t,x)\approx 0$}
\put(-145,200){$C(t,x)\sim \mathcal{O}(1)$}

\end{center}
\caption{\small Light cone (gray region) and butterfly effect cone (dark gray region). Inside the butterfly effect cone, for $t-t_* \geq |x|/v_B$, we have $C(t,x) \sim \mathcal{O}(1)$, whereas for outside the cone, for $t-t_* < |x|/v_B$, the commutator is small, $C(t,x) \sim 1/N_\mt{dof} <<1$. Outside the light-cone the Lorentz invariance implies a zero commutator.}
\label{fig-VBcone}
\end{figure}

In quantum field theories we can upgrade (\ref{eq-C(t)}) to the case where the operators are separated in space
\be
C(t,x) = \langle -[V(0,0), W(t,x)]^2 \rangle_{\beta}\,.
\label{eq-C(t,x)}
\ee
Strictly speaking, the above expression is generically divergent, but it can be regularized by adding imaginary times to the time arguments of the operators $V$ and $W$.  For  a large class of spin-chains, higher dimensional SYK-models and CFTs the above commutator is roughly given by
\be
C(t,x) \sim \exp \left[\lambda_L \left(t-t_*-\frac{|x|}{v_B} \right) \right]\,,
\ee
where $v_B$ is the so-called butterfly velocity\footnote{Actually, $v_B$ represents the ``velocity of the butterfly effect''. Here we continue to follow the tradition of using misnomers.}. This velocity describes the growth of the operator $W$ in physical space and it acts as a low-energy Lieb-Robinson velocity \cite{Roberts:2016wdl}, which sets a bound for the rate of transfer of quantum information. From the above formula, we can see that there is an additional delay in scrambling due to the physical separation between the operators. The butterfly velocity defines an effective light-cone for the commutator (\ref{eq-C(t,x)}). Inside the cone, for $t-t_* \geq |x|/v_B$, we have $C(t,x) \sim \mathcal{O}(1)$, whereas for outside the cone, for $t-t_* < |x|/v_B$, the commutator is small, $C(t,x) \sim 1/N_\mt{dof} <<1$. Outside the light-cone the Lorentz invariance implies a zero commutator. The light-cone and the butterfly effect cone are illustrated in figure \ref{fig-VBcone}.

\section{Chaos $\&$ Holography}\label{sec-4}

In this section, we review how the chaotic properties of holographic theories can be described in terms of black holes physics.
Black holes behave as thermal systems and thermal systems generically display chaos. This implies that black holes are somehow chaotic. This statement has a precise realization in the context of the gauge/gravity duality. According to this duality, some strongly coupled non-gravitational systems are dual to higher dimensional gravitational systems. In the most known and studied example of this duality the $\mathcal{N}=4$ super Yang-Mills (SYM) theory living in $R^{3,1}$ is dual to type IIB supergravity in $AdS_5 \times S^5$. More generically, a $d-$dimensional non-gravitational theory living in $R^{d-1,1}$ is dual to a  gravity theory living in a higher-dimensional space of the form $AdS_{d+1} \times \mathcal{M}$, where $\mathcal{M}$ is generically a compact manifold. The non-gravitational theory can be thought as living in the boundary of $AdS_{d+1}$ and because of that is usually called the {\it boundary theory}. The gravitational theory is also called the {\it bulk theory}. 

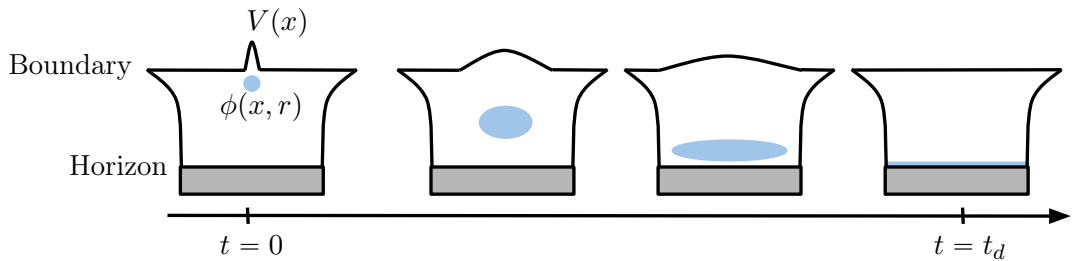
\begin{figure}[H]
\begin{center}
\includegraphics[width=12.5cm]{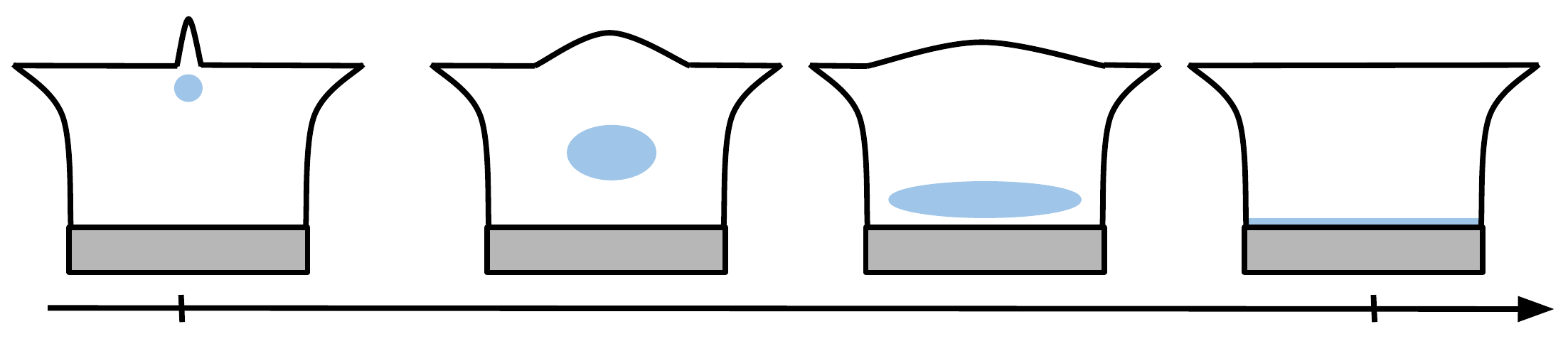}
\put(-315,80){\small $V(x)$}
\put(-325,48){\small $\phi(x,r)$}
\put(-325,-5){\small $t=0$}
\put(-55,-5){\small $t=t_d$}
\put(-405,63){\small Boundary}
\put(-382,25){\small Horizon}

\end{center}
\caption{\small Bulk picture of thermalization. The figure represents an asymptotically $AdS$ black hole geometry. The boundary is at the top edge, while the black hole horizon is at the bottom edge. The black hole's interior is shown in gray. The boundary operator $V$ is dual to the bulk field $\phi$. From the point of view of the boundary theory the perturbation produced by $V$ is initially localized around the point $x$, but it gets delocalized over a region that increases with time. In the bulk description this is described by a particle (field excitation) that is initially close to the boundary and then falls into the black hole.}
\label{fig-thermalization}
\end{figure}

There is a dictionary relating physical quantities in the boundary and bulk description \cite{duality2,duality3}. An example is provided by the operators of the boundary theory, which are related to bulk fields. The boundary theory at finite temperature can be described by introducing a black hole in the bulk. The thermalization properties of the boundary theory have a nice visualization in terms of black holes physics. By applying a local operator in the boundary theory we produce some perturbation that describes a small deviation from the thermal equilibrium. The information about $V(x)$ is initially contained around the point $x$, but it gets delocalized over a region that increases with time until it completely melts into the thermal bath. In the bulk theory, the application of the operator $V(x)$ produces a particle (field excitation) close to the boundary of the space, which then falls into the black hole. The return to the thermal equilibrium in the boundary theory corresponds to the absorption of the bulk particle by the black hole.
Figure \ref{fig-thermalization} illustrates the bulk description of thermalization.

The approach to thermal equilibrium is controlled by the black hole's quasi-normal modes (QNMs). In holographic theories, the quasi-normal modes control the decay of two-point functions of the boundary theory

\be
\langle V(t)V(0)\rangle_{\beta} \sim e^{-t/t_d}
\ee
where the dissipation time $t_d$ is related to the lowest quasi-normal mode ($\text{Im}(\omega) \sim t_d^{-1}$). From the point of view of the bulk theory the QNMs describes how fast a perturbed black hole returns to equilibrium.
Clearly, the black hole's quasi-normal modes correspond to the classical Ruelle resonances. In holographic theories the dissipation time is roughly give by $t_d \sim \beta$.

Another important exponential behavior of black holes is provided by the blueshift suffered by the in-falling quanta, or, equivalently, the red shift suffered by the quanta escaping from the black hole. The blueshift suffered by the in-falling quanta is determined by the black hole's temperature. If the quanta asymptotic energy is $E_0$, this energy increases exponentially with time
\be
E=E_0\, e^{\frac{2\pi}{\beta}t}\,,
\ee
where $\beta$ is the Hawking's inverse temperature. Later we will see that this exponential increase in the energy of the in-falling quanta gives rise to the Lyapunov behavior of $C(t,x)$ of holographic theories.

\subsection{Holographic setup}
\subsubsection*{The TFD state $\&$ Two-sided black holes}\label{sec-TFD}

In the study of chaos is convenient to consider a thermofield double state made out of two identical copies of the boundary theory
\be
|\text{TFD} \rangle = \frac{1}{Z^{1/2}} \sum_n e^{-\beta E_n /2} | n \rangle_\mt{L} |n \rangle_\mt{R}\,,
\ee
where $L$ and $R$ label the states of the two copies, which we call $\text{QFT}_\mt{L}$ and $\text{QFT}_\mt{R}$, respectively. The two boundary theories do not interact and only know about each other through their entanglement. This state is dual to an eternal (two-sided) black hole, with two asymptotic boundaries, where the boundary theories live \cite{eternalBH}. This is a wormhole geometry, with an Einstein-Rosen bridge connecting the two sides of the geometry. The wormhole is not traversable, which is consistent with the fact that the two boundary theories do not interact.

For definiteness we assume a metric of the form
\be
ds^2 = -G_{tt}(r)dt^2+G_{rr}(r)dr^2+G_{ij}(r,x^k)dx^i dx^j\,,
\label{eq-metric0}
\ee
where the boundary is located at $r=\infty$, where the above metric is assumed to asymptote $AdS_{d+1}$. We take the horizon as located at $r=\rh$, where $G_{tt}$ vanish and $G_{rr}$ has a first order pole. For future purposes, let $\beta$ be the Hawking's inverse temperature, and $S_\mt{BH}$ be the Bekenstein-Hawking entropy.

\begin{figure}[H]
\centering
%\captionsetup{justification=centering}

\begin{tikzpicture}[scale=1.5]
\draw [thick]  (0,0) -- (0,3);
\draw [thick]  (3,0) -- (3,3);
\draw [thick,dashed]  (0,0) -- (3,3);
\draw [thick,dashed]  (0,3) -- (3,0);
\draw [thick,decorate,decoration={zigzag,segment length=1.5mm, amplitude=0.3mm}] (0,3) .. controls (.75,2.85) 
and (2.25,2.85) .. (3,3);
\draw [thick,decorate,decoration={zigzag,segment length=1.5mm,amplitude=.3mm}]  (0,0) .. controls (.75,.15) and (2.25,.15) .. (3,0);

\draw[thick,<->] (1,2.2) -- (1.5,1.7) -- (2,2.2);

\node[scale=0.8, align=center] at (1.5,2.65) {Future Interior};
\node[scale=0.8,align=center] at (1.5,.55) {Past Interior};
\node[scale=0.8,align=center] at (0.6,1.6) {Left\\ Exterior};
\node[scale=0.8,align=center] at (2.4,1.6) {Right\\ Exterior};
\end{tikzpicture}
\put(10,60){\Large $= |TFD \rangle$}
\vspace{0.1cm}
\put(-137,25){\rotatebox{90}{\small $r = \infty$}}
\put(3,25){\rotatebox{90}{\small $r = \infty$}}
\put(-75,-5){\small $r = 0$}
\put(-75,130){\small $r = 0$}
\put(-92,95){\small $U$}
\put(-43,95){\small $V$}

\caption{ \small Penrose diagram for the two-sided black holes with two boundaries that asymptote $AdS$. This geometry is dual to a thermofield double state constructed out of two copies of the boundary theory.}
\label{fig-Penrose}
\end{figure}
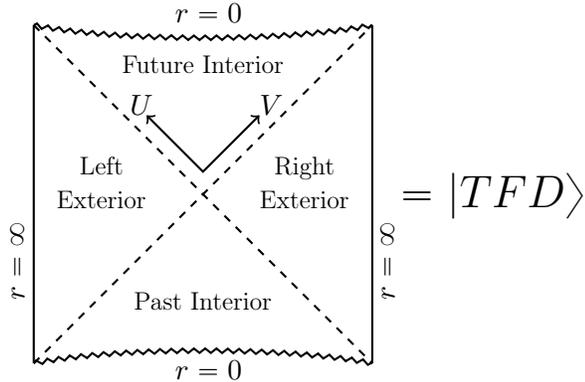

In the study of shock waves is more convenient to work with Kruskal-Szekeres coordinates, since these coordinate cover smoothly the globally extended spacetime. We first define the tortoise coordinate
\begin{equation}
d r_*=\sqrt{\frac{G_{rr}}{G_{tt}}} dr\,,
\end{equation}
and then we introduce the Kruskal-Szekeres coordinates $U,V$ as
\bea
U&=&+e^{\frac{2\pi}{\beta}\left(r_*-t\right)}\,,\,\,V=-e^{\frac{2\pi}{\beta}\left(r_*+t\right)} \,\,\,(\text{left exterior region}) \nonumber \\
U&=&-e^{\frac{2\pi}{\beta}\left(r_*-t\right)}\,,\,\,V=+e^{\frac{2\pi}{\beta}\left(r_*+t\right)} \,\,\,(\text{right exterior region})  \\
U&=&+e^{\frac{2\pi}{\beta}\left(r_*-t\right)}\,,\,\,V=+e^{\frac{2\pi}{\beta}\left(r_*+t\right)} \,\,\,(\text{future interior region}) \nonumber \\
U&=&-e^{\frac{2\pi}{\beta}\left(r_*-t\right)}\,,\,\,V=-e^{\frac{2\pi}{\beta}\left(r_*+t\right)} \,\,\,(\text{past interior region}) \nonumber 
\eea
In terms of these coordinates the metric reads
\begin{equation}
ds^2=2A(UV)dU dV+G_{ij}(UV)dx^i dx^j\,,
\end{equation}
where
\begin{equation}
A(UV)=\frac{\beta^2}{8\pi^2}\frac{G_{tt}(UV)}{UV}\,.
\label{eq-metric0Kruskal}
\end{equation}
In these coordinates the horizon is located at $U=0$ or at $V=0$. The left and right boundaries are located at $UV=-1$ and the past and future singularities at $UV=1$. The Penrose diagram for this metric is shown in figure \ref{fig-Penrose}.

The global extended spacetime can also be described in terms of complexified coordinates \cite{Fidkowski:2003nf}. In this case one defines the complexified Schwarzschild time
\be
t=t_\mt{L}+i\,t_\mt{E}\,,
\ee
where $t_\mt{L}$ and $t_\mt{E}$ are the Lorentzian and Euclidean times, and then one describes the time in each of the four patches (left and right exterior regions, and the future and past interior regions) as having a constant imaginary part
\bea
t_\mt{E}&=&0  \,\,\,\,\,\,\,(\text{right exterior region}) \nonumber \\
t_\mt{E}&=&-\beta/4 \,\,\,(\text{future interior region})\\
t_\mt{E}&=&-\beta/2  \,\,\,(\text{left exterior region}) \nonumber \\
t_\mt{E}&=&+ \beta/4 \,\,\,(\text{past interior region}) \nonumber
\eea
The Euclidean time has a period of $\beta$. The Lorentzian time increases upward (downward) in the right (left) exterior region, and to the right (left) in the future (past) interior.

Note that, with the complexified time, one can obtain an operator acting on the left boundary theory by adding (or subtracting) $i \beta/2$ to the time of an operator acting on the right boundary theory.

\subsubsection*{Perturbations of the TFD state $\&$ Shock wave geometries}

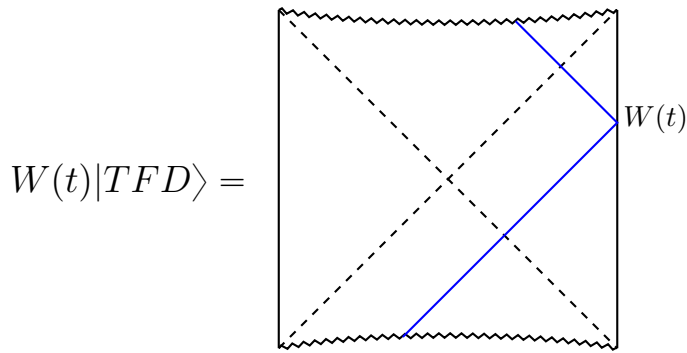
\begin{figure}[H]
\begin{center}
\begin{tikzpicture}[scale=1.5]
\draw [thick]  (0,0) -- (0,3);
\draw [thick]  (3,0) -- (3,3);
\draw [thick,dashed]  (0,0) -- (3,3);
\draw [thick,dashed]  (0,3) -- (3,0);
\draw [thick,decorate,decoration={zigzag,segment length=1.5mm, amplitude=0.3mm}] (0,3) .. controls (.75,2.85) 
and (2.25,2.85) .. (3,3);
\draw [thick,decorate,decoration={zigzag,segment length=1.5mm,amplitude=.3mm}]  (0,0) .. controls (.75,.15) and (2.25,.15) .. (3,0);

\draw [thick,blue]  (1.1,0.1) -- (3,2) -- (2.1,2.9);

\end{tikzpicture}
\put(2,85){\small $W(t)$}
\put(-230,60){\large $W(t)|TFD \rangle =$}
\end{center}
\caption{\small Bulk description of the state $W(t)|\text{TFD}\rangle$. In blue is shown the trajectory of a `particle' that comes out of the past horizon, reaches the boundary at time $t$ producing the perturbation $W$, and then falls into the future horizon. For now on, we will refer to this bulk excitation as the {\it W-particle}.}
\label{fig-W-state}
\end{figure}

We now turn to the description of states of the form 
\be
W(t) |\text{TFD} \rangle
\ee
where $W$ is a thermal scale operator that acts on the right boundary theory. This state can be describe by a `particle' (field excitation) in the bulk that comes out of the past horizon, reaches the right boundary at time $t$, and then falls into the future horizon, as illustrated in figure \ref{fig-W-state}.

If $|t|$ is not too large, the state $W(t) |\text{TFD}\rangle$ will represent just a small perturbation of the TFD state and the corresponding description in the bulk will be just an eternal two-sided black hole geometry slightly perturbed by the presence of a probe particle. This is no longer the case if $|t|$ is large. In this case, there is a non-trivial modification of the geometry. A very early perturbation, for example, is described in the bulk in terms of a particle that falls towards the future horizon for a very long time and gets highly blueshifted in the process. If the particle's energy is $E_0$ in the asymptotic past, this energy will be exponentially larger from the point of view of the $t=0$ slice of the geometry, i.e., $E=E_0\,e^{\frac{2\pi}{\beta}t}$. Therefore, for large enough $|t|$, the particle's energy will be very large and one needs to include the corresponding back-reaction.
%
%The same is true for a very late perturbation. Because of the gravitational red-shift, the W-particle can only reach the boundary (at some late time) with thermal energy if it were travelling near the future horizon with a very high energy.

\begin{figure}[H]
\begin{center}
\begin{tikzpicture}[scale=1.5]
\draw [thick]  (0,0) -- (0,3);
\draw [thick]  (3,0) -- (3,3);
\draw [thick,dashed]  (0,0) -- (3,3);
\draw [thick,dashed]  (0,3) -- (3,0);
\draw [thick,decorate,decoration={zigzag,segment length=1.5mm, amplitude=0.3mm}] (0,3) .. controls (.75,2.85) 
and (2.25,2.85) .. (3,3);
\draw [thick,decorate,decoration={zigzag,segment length=1.5mm,amplitude=.3mm}]  (0,0) .. controls (.75,.15) and (2.25,.15) .. (3,0);

\draw [thick,blue]  (3,0.1) -- (0.1,3.);

\draw [thick,->,red]  (1.5,.5) -- (1.9,.9);
\draw [thick,red]  (1.5,.5) -- (2.06,1.06);

\draw [thick,->,red]  (1.66,1.46) -- (2.06,1.86);

%\draw [thick,decorate,decoration={zigzag,segment length=1.5mm, amplitude=0.3mm}]  (4.,0) --(7.,0);
%\draw [thick,decorate,decoration={zigzag,segment length=1.5mm, amplitude=0.3mm}]  (5,3) --(8,3);
%\draw [thick]  (4.,0) --(5,3);
%\draw [thick]  (7.,0) --(8,3); 
%\draw [thick] (7,0) -- (5,3);
%\draw [thick] (7.05,0.05) -- (5.05,3.05);
%\draw [thick,dashed] (4,0) -- (5.7,1.7);
%\draw [thick,dashed] (8,3) -- (6.3,1.3);

\end{tikzpicture}
\put(2,3){\small $W(-t)$}
\put(-48,55){\small $h(t,\vec x)$}
\put(-230,60){\large $W(-t)|TFD \rangle =$}
%\put(-150,60){\large or}
\end{center}
\caption{\small Bulk description of the state $W(-t)|\text{TFD}\rangle$. An early enough perturbation produces a shock wave geometry. The effect of the shock wave (shown in blue) is to produce a shift $U \rightarrow U+h(t, \vec x)$ in the trajectory of a probe particle (shown in red) crossing it.}
\label{fig-W-state2}
\end{figure}
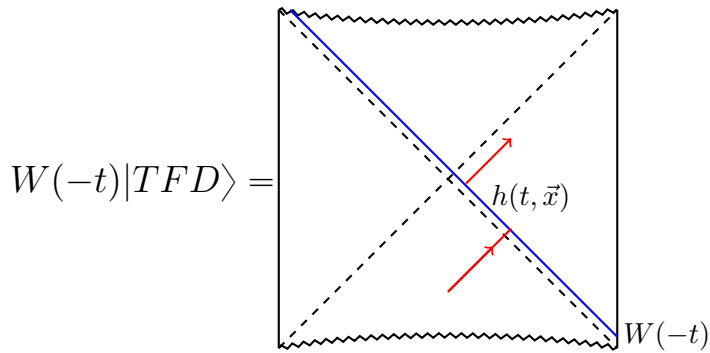

The back-reaction of a very early (or very late) perturbation is actually very simple - it corresponds to a shock wave geometry \cite{Dray-85,Sfetsos-94}. To understand that, we first need to notice that, under boundary time evolution, the stress-energy of a generic perturbation $W$ gets compressed in the $V-$direction, and stretched in the $U-$direction. For large enough $|t|$ we can approximate the stress tensor of the W-particle as
\be
T_{VV} \sim P^{U} \delta (V) a(\vec x)\,,
\label{eq-stressVV}
\ee
where $P^{U} \sim \beta^{-1} e^{\frac{2\pi}{\beta}t}$ is the momentum of the W-particle in the $U-$direction and $a(\vec x)$ is some generic function that specifies the location of the perturbation in the spatial directions of the right boundary. Note that $T_{VV}$ is completely localized at $V=0$ and homogeneous along the $U-$direction. Besides that, even if the W-particle is massive, the exponential blue-shift will make it follow an almost null trajectory, as shown in figure \ref{fig-W-state2}.

The shock wave geometry produced by the W-particle is described by the metric
\be
ds^2=2A(UV)dU dV+G_{ij}(UV)dx^i dx^j-2A(UV)h(t,\vec{x})\delta(V) dV^2\,,
\ee
that is completely specified by the shock wave transverse profile $h(t,\vec x)$. This geometry can be seen as two pieces of a eternal black hole glued together along $V=0$ with a shift of magnitude $h(t,\vec x)$ in the $U-$direction. We find it useful to represent this geometry with the same Penrose diagram of the unperturbed geometry, but with the prescription that any trajectory crossing the shock wave gets shifted in the $U-$direction as $U \rightarrow U +h(t,\vec x)$. See figure \ref{fig-W-state2}. 

The precise form of $h(t,\vec x)$ can be determined by solving the $VV-$component of Einstein's equation. For a local perturbation, i.e., $a(\vec x)=\delta^{d-1}(\vec x)$, the solution reads

\be
h(t,\vec{x}) \sim G_\mt{N}\,e^{\frac{2\pi}{\beta}t-\mu |\vec{x}| }\,,\,\,\,\,\text{with}\,\,
\mu=\frac{2\pi}{\beta} \sqrt{\frac{(d-1)G'_{ii}(\rh)}{G'_{tt}(\rh)}}\,,
\ee
where, for simplicity, $G_{ij}$ has been assumed to be diagonal and isotropic.

Interestingly, the shock wave profile contains information about the parameters characterizing the chaotic behavior of the boundary theory. Indeed, the double commutator has a region of exponential growth at which $C(t,\vec x) \sim h(t,\vec x)$. From this identification, we can write
\be
h(t,\vec x) \sim e^{\frac{2\pi}{\beta}\left( t-t_*-\frac{|\vec x|}{v_B} \right)}
\ee
where (the leading order contribution) to the scrambling time scales logarithmically with the Bekenstein-Hawking entropy
\be
t_* \sim \frac{\beta}{2\pi}\log \frac{1}{G_\mt{N}}\sim\frac{\beta}{2\pi}\log S_\mt{BH}\,,
\ee
while the Lyapunov exponent is proportional to the Hawking's temperature.
\be
\lambda_L=\frac{2\pi}{\beta}\,.
\ee
The butterfly velocity is determined from the near-horizon geometry\footnote{Here we are assuming isotropy. In the case of anisotropic metrics the formula for $v_B$ is a little bit more complicated. See, for instance, the appendix A of \cite{Fischler:2018kwt} or the appendix B of \cite{Baggioli-2018}.}
\be
v_{B}^2=\frac{G_{tt}'(\rh)}{(d-1)G_{ii}'(\rh)}\,.
\ee

\subsection{Bulk picture for the behavior of OTOCs} \label{sec-bulkpictureOTOC}

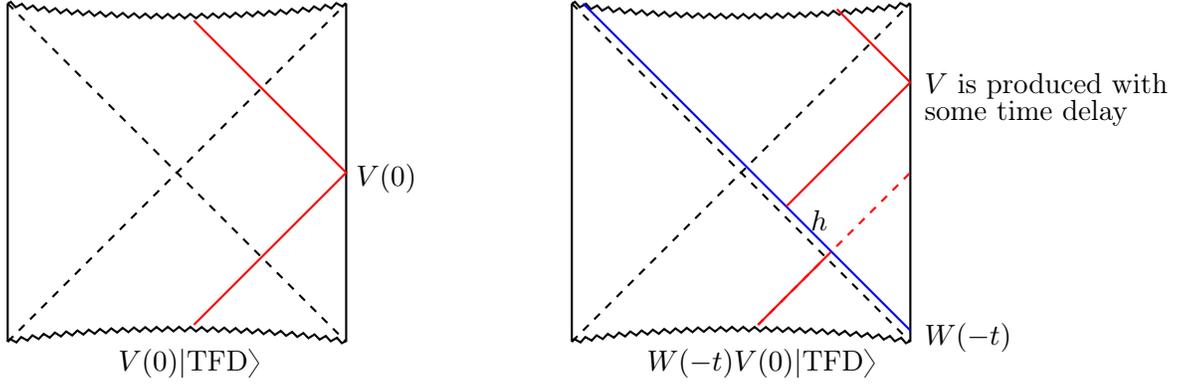
\begin{figure}[H]
\begin{center}

\begin{tikzpicture}[scale=1.5]
\draw [thick]  (0,0) -- (0,3);
\draw [thick]  (3,0) -- (3,3);
\draw [thick,dashed]  (0,0) -- (3,3);
\draw [thick,dashed]  (0,3) -- (3,0);
\draw [thick,decorate,decoration={zigzag,segment length=1.5mm, amplitude=0.3mm}] (0,3) .. controls (.75,2.85) 
and (2.25,2.85) .. (3,3);
\draw [thick,decorate,decoration={zigzag,segment length=1.5mm,amplitude=.3mm}]  (0,0) .. controls (.75,.15) and (2.25,.15) .. (3,0);

\draw [thick,red]  (1.65,0.15) -- (3,1.5) -- (1.65,2.85);

\draw [thick]  (5,0) -- (5,3);
\draw [thick]  (8,0) -- (8,3);
\draw [thick,dashed]  (5,0) -- (8,3);
\draw [thick,dashed]  (5,3) -- (8,0);
\draw [thick,decorate,decoration={zigzag,segment length=1.5mm, amplitude=0.3mm}] (5,3) .. controls (5.75,2.85) 
and (7.25,2.85) .. (8,3);
\draw [thick,decorate,decoration={zigzag,segment length=1.5mm,amplitude=.3mm}]  (5,0) .. controls (5.75,.15) and (7.25,.15) .. (8,0);

\draw [thick,blue]  (8,0.1) -- (5.1,3.);

\draw [red,thick] (6.65,0.15) -- (7.25,.75);

\draw [red,thick] (6.9,1.2) -- (8.,2.3);

\draw [red,thick] (8,2.3) -- (7.35,2.95);

\draw [red,thick,dashed] (6.65,0.15) -- (8,1.5);

\end{tikzpicture}
\put(5,0){\small $W(-t)$}
\put(5,95){\small $V$ is produced with}
\put(5,85){\small some time delay}
\put(-210,60){\small $V(0)$}
\put(-300,-10){\small $V(0)| \text{TFD} \rangle$}
\put(-100,-10){\small $W(-t)V(0)| \text{TFD} \rangle$}
\put(-38,43){\small $h$}

\end{center}
\caption{ \small {\it Left panel}: bulk description of the  state $ V(0)| \text{TFD} \rangle$. The V-particle comes out of the past horizon, reaches the boundary at time $t=0$ producing the perturbation $V$, and then falls into the future horizon. {\it Right panel}: bulk description of the `in' state $| \psi_\mt{in} \rangle = W(-t)V(0)| \text{TFD} \rangle$. The W-particle (shown in blue) produces a shock wave along $V=0$. The trajectory of the V-particle (shown in red) suffers a shift and the perturbation $V$ is produced at the boundary with some time delay.}
\label{fig-V-state}
\end{figure}

In this section we present the bulk perspective for the vanishing of OTOCs at later times. In order to do that, we write the OTOC as a superposition of two states 
\be
\text{OTO}(t) = \langle \text{TFD} | W(-t)V(0)W(-t)V(0) |\text{TFD} \rangle = \langle \psi_\mt{out} | \psi_\mt{in} \rangle\,,
\ee
where the `in' and `out' states are given by
\be
|\psi_\mt{in} \rangle =W(-t)V(0)| \text{TFD} \rangle \,, \,\,\,\, |\psi_\mt{out} \rangle =V^{\dagger}(0)W^{\dagger}(-t)| \text{TFD} \rangle
\ee
The interpretation of a vanishing OTOC in terms of the bulk theory is actually very simple. Let us go step by step and construct first the state $V(0)| \beta \rangle$. This state is described by a particle that comes out of the past horizon, reaches the boundary at $t=0$, and then falls back into the future horizon. See the left panel of figure \ref{fig-V-state}.

Now the `in' state can be obtained as
\be
|\psi_\mt{in} \rangle =W(-t)V(0)|\text{TFD} \rangle\, = e^{-iHt}\,W(0)\,e^{iHt}V(0)|\text{TFD} \rangle\,.
\ee
This amounts to evolve the state $V(0)|\text{TFD} \rangle$ backwards in time, apply the operator $W$, and then evolve the system forwards in time. The corresponding description in the bulk is shown in the right panel of figure \ref{fig-V-state}. From this picture we can see that the perturbation $W$ produces a shock wave that causes a shift in the trajectory of the V-particle, which no longer reaches the boundary at time $t=0$, but rather with some time delay. The physical interpretation is that a small perturbation  in the asymptotic past (represented by $W$) is amplified over time and destroys the initial configuration (represented by the state $V(0)|\text{TFD} \rangle$).

\begin{figure}[H]
\begin{center}
\begin{tikzpicture}[scale=1.5]
\draw [thick]  (5,0) -- (5,3);
\draw [thick]  (8,0) -- (8,3);
\draw [thick,dashed]  (5,0) -- (8,3);
\draw [thick,dashed]  (5,3) -- (8,0);
\draw [thick,decorate,decoration={zigzag,segment length=1.5mm, amplitude=0.3mm}] (5,3) .. controls (5.75,2.85) 
and (7.25,2.85) .. (8,3);
\draw [thick,decorate,decoration={zigzag,segment length=1.5mm,amplitude=.3mm}]  (5,0) .. controls (5.75,.15) and (7.25,.15) .. (8,0);

\draw [thick,blue]  (8,0.1) -- (5.1,3.);
%T\draw [red,thick] (6.65,0.15) -- (7.25,.75);

\draw [red,thick] (7.3,0.8) -- (8,1.5);

\draw [red,thick] (7.7,0.4) -- (7.35,0.05);

\draw [red,thick] (8,1.5) -- (6.6,2.9);

\end{tikzpicture}
\put(-100,-10){\small $V(0)W(-t)| \text{TFD} \rangle$}
\put(3,0){\small $W(-t)$}
\put(3,60){\small $V(0)$}
\put(-20,26){\small $h$}

\end{center}

\caption{\small Bulk description of the `out' state $|\psi_\mt{out} \rangle = V(0)W(-t)| \text{TFD} \rangle$. The W-particle produces the shock wave geometry. The trajectory of the V-particle is such that, after suffering the shift $U \rightarrow U + h(t,\vec x)$, it reaches the boundary at time $t=0$, producing the perturbation $V$.}
\label{fig-VWstate}
\end{figure}
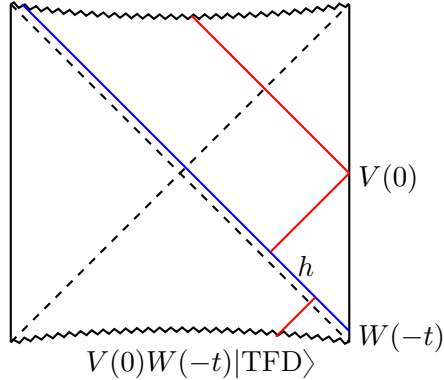

The bulk description of the `out' state can be obtained in the same way. As this state displays the perturbation $V$ at $t=0$, the V-particle should be produced in the asymptotic past in such a way that, after its trajectory gets shifted as $U \rightarrow U + h$, it reaches the boundary at the time $t=0$ producing the perturbation $V$.

Comparing the bulk description of the state $| \psi_\mt{in}\rangle$ (shown in the right panel of figure \ref{fig-V-state}) with the description of the state $|\psi_\mt{out}\rangle$ (shown in figure \ref{fig-VWstate}) we can see that these states are indistinguishable when $h(t,\vec x)$ is zero, but they become more and more different for large values of $h(t,\vec x)$. As a consequence, the overlap $C(t)= \langle \psi_\mt{out} | \psi_\mt{in} \rangle$ is equal to one when $h=0$, but it decreases to zero as we increase the value of $h$.

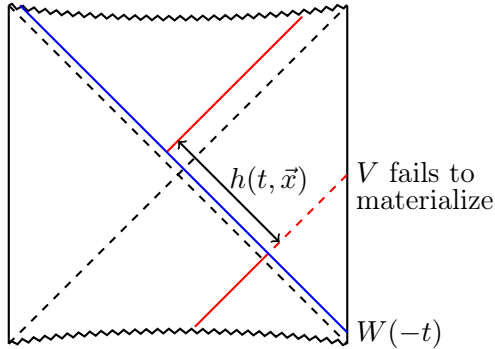
\begin{figure}[H]
\begin{center}
\begin{tikzpicture}[scale=1.5]
\draw [thick]  (0,0) -- (0,3);
\draw [thick]  (3,0) -- (3,3);
\draw [thick,dashed]  (0,0) -- (3,3);
\draw [thick,dashed]  (0,3) -- (3,0);
\draw [thick,decorate,decoration={zigzag,segment length=1.5mm, amplitude=0.3mm}] (0,3) .. controls (.75,2.85) 
and (2.25,2.85) .. (3,3);
\draw [thick,decorate,decoration={zigzag,segment length=1.5mm,amplitude=.3mm}]  (0,0) .. controls (.75,.15) and (2.25,.15) .. (3,0);

\draw [thick,red]  (1.65,0.15) -- (2.3,0.8);
\draw [thick,red,dashed]  (2.35,0.85) -- (3,1.5);

\draw [thick,red]  (1.4,1.7) -- (2.6,2.9);

\draw [thick,blue]  (3,0.1) -- (0.1,3.);

\draw [thick,<->]  (1.5,1.8) -- (2.4,.9);

\end{tikzpicture}
\put(3,2){\small $W(-t)$}
\put(-45,60){\small $h(t,\vec x)$}
\put(3,62){\small $V$ fails to}
\put(3,52){\small materialize}

\end{center}
\caption{\small  Bulk description of the state $W(-t)V(0)|\text{TFD}\rangle$ for the case where $|t| \gtrsim t_*$. The V-particle's trajectory undergoes a shift, and it is captured by the black hole. The perturbation $V$ never forms, and the corresponding state have no superposition with the `out' state $V(0)W(-t)|\text{TFD}\rangle$, resulting in a vanishing OTOC.}
\label{fig-zeroC}
\end{figure}

%The precise behaviour of the double commutator $C(t)$ depends on the shock wave profile $h(t,\vec x)$ and on the operators\footnote{In the case where the boundary theory is a CFT, the double commutator depends on the shock wave profile and on the scaling dimensions of the operator $V$ and $W$.} $V$ and $W$. There is, however, a universal behaviour for some range of time at which the commutator is basically proportional to the shock wave profile, i.e.,
%\be
%C(t, \vec x) \sim h(t,\vec x) = \exp \left[\frac{2\pi}{\beta} \left(t-t_*-\frac{|\vec x|}{v_B} \right) \right]\,.
%\ee
The exponential behavior of $h(t,\vec x)$ implies that an early enough perturbation can produce a very large shift in the V-particle's trajectory, causing it to be captured by the black hole, and preventing the materialization of the $V$ perturbation at the boundary. See figure \ref{fig-zeroC}. This should be compared with the physical picture given in figure \ref{fig-WVstate1}.

\begin{figure}[H]
\begin{center}
\begin{tabular}{c@{\hskip 1in}c}
\setlength{\unitlength}{1cm}
\hspace{-0.9cm}
\includegraphics[width=6.5cm]{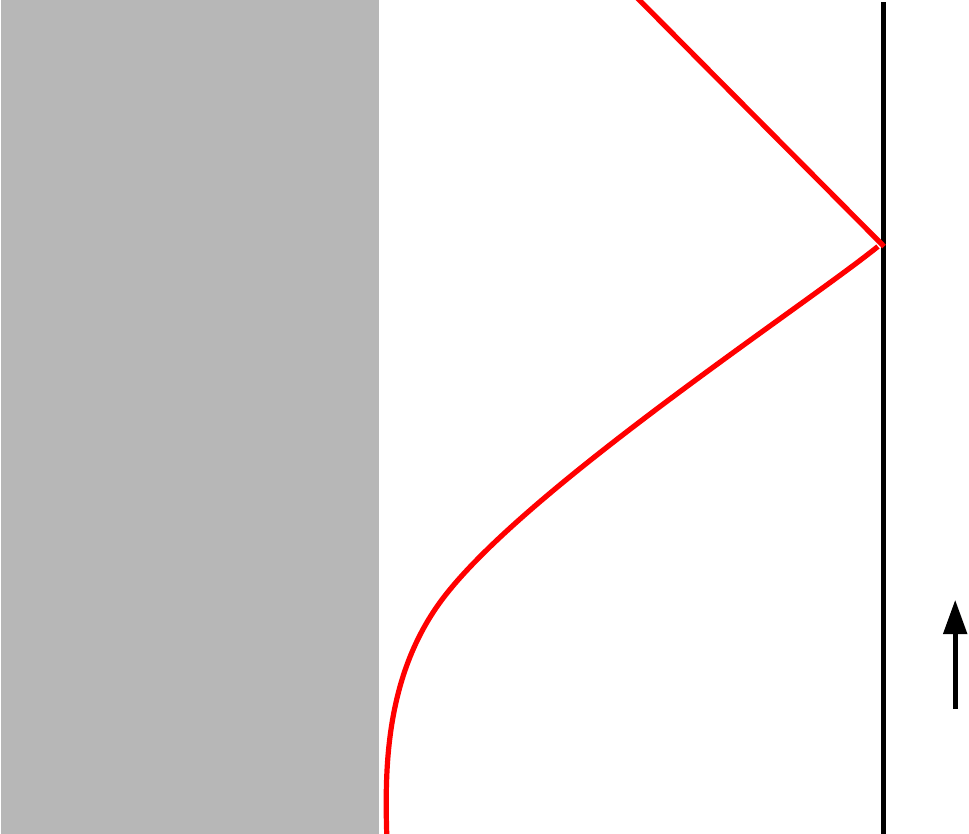}
\qquad\qquad &
\includegraphics[width=6.5cm]{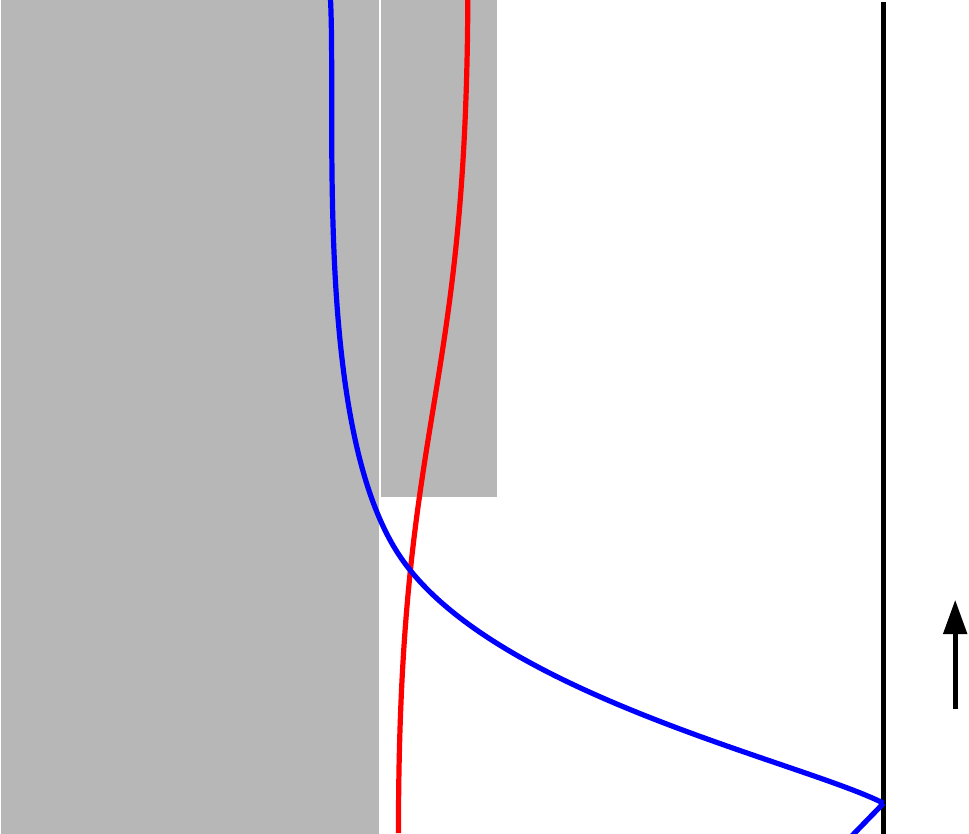}
\qquad

\end{tabular}
\put(-310,15){\rotatebox{-90}{\small Boundary}}
\put(-405,15){\rotatebox{-90}{\small Horizon}}
\put(-120,80){\rotatebox{-90}{\small Old Horizon}}
\put(-98,80){\rotatebox{-90}{\small New Horizon}}
%\put(-250,35){\small $V(0)$ fails to}
%\put(-250,23){\small re-materialize}
%\put(-250,-70){\small $W(-t)$}
\put(-308,35){\small $V(0)$}
\put(-290,-45){\small $t$}
\put(-5,-45){\small $t$}
\put(-310,15){\rotatebox{-90}{\small Boundary}}
\put(-25,15){\rotatebox{-90}{\small Boundary}}
%\put(-348,80){\rotatebox{-90}{\small Old Horizon}}
%\put(-326,80){\rotatebox{-90}{\small New Horizon}}
\put(-420,-85){\small $r=\rh$}
\put(-320,-85){\small $r=\infty$}
\put(-22,35){\small $V(0)$ fails to }
\put(-22,25){\small materialize}
%\put(-55,-12){\small $\alpha$}
\put(-22,-72){\small $W(-t)$}

\end{center}

\caption{ \small {\it Left panel}: bulk description of the state  $V(0)|\text{TFD}\rangle$. The V-particle (whose trajectory is shown in red) escapes from the black hole (shown in gray), reaches the boundary at time $t=0$, and then falls back towards the horizon. {\it Right panel}: bulk description of the state $W(-t)V(0)|\text{TFD}\rangle$ for the case where $|t| \gtrsim t_*$. The W-particle (whose trajectory is shown in blue) gets blue shifted and increases the black hole size as it falls into it. The V-particle fails to scape from the larger black hole, and the perturbation $V$ never forms at the boundary.}
\label{fig-WVstateHeuristic}
\end{figure}
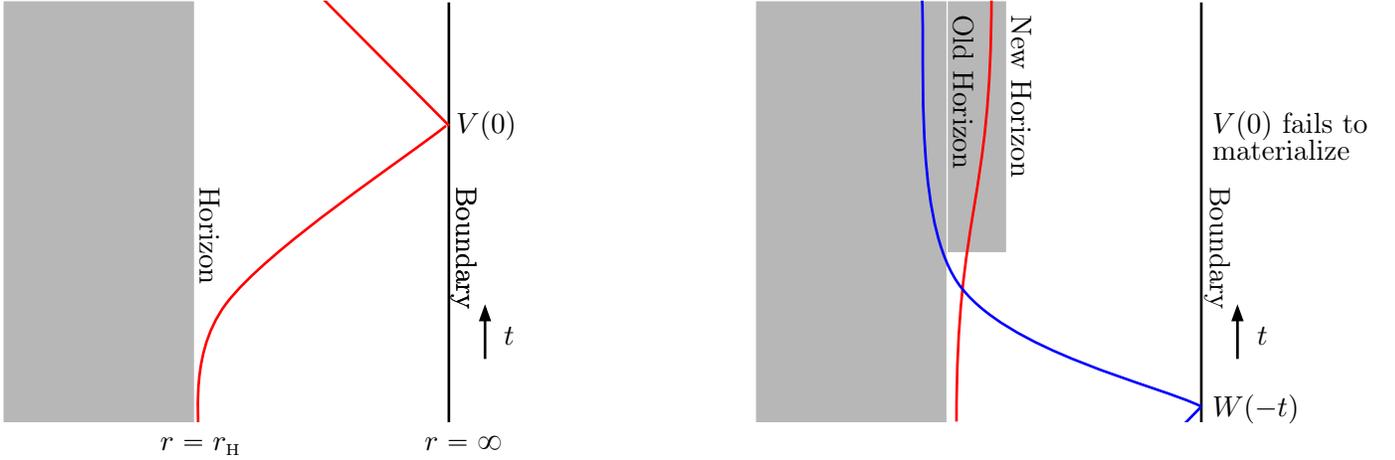

The physical picture of the process described in figure \ref{fig-zeroC} is quite simple. The state $V(0)| \text{TFD}\rangle$ can be represented by a black hole geometry in which a particle (the V-particle) escapes from the black holes and reaches the boundary at time $t=0$. The state $W(-t)V(0)| \text{TFD}\rangle$ is obtained by perturbing the state $V(0)| \text{TFD}\rangle$ in the asymptotic past. This corresponds to add a W-particle to the system in the asymptotic past. This particle gets highly blue shifted as it falls towards the black hole. The black hole captures the W-particle and becomes bigger. The V-particle fails to escape from the bigger black hole, and never reaches the boundary to produce the $V$ perturbation. This physical picture is illustrated in figure \ref{fig-WVstateHeuristic}.

The precise form of the above OTOC can be obtained by calculating the overlap $\langle \psi_\mt{out}|\psi_\mt{in}\rangle$ using the Eikonal approximation \cite{BHchaos4}, in which the Eikonal phase $\delta$ is proportional to the shock wave profile $\delta \sim h(t,\vec x)$. The OTOC can be written as an integral of the phase $e^{i\delta}$ weighted by kinematical factors which are basically Fourier transforms of bulk-to-boundary propagators for the $V$ and $W$ operators.

The result for Rindler AdS$_3$ reads\footnote{The above result assumes $\Delta_W >> \Delta_V$. }
\be
\frac{\langle V(i\epsilon_1) W(t+i\epsilon_2) V(i\epsilon_3) W(t+i\epsilon_4) \rangle}{\langle V(i\epsilon_1) V(i\epsilon_3) \rangle \langle W(i\epsilon_2) W(i\epsilon_4) \rangle}= \left( \frac{1}{1-\frac{8 \pi i G_\mt{N} \Delta_W}{\epsilon_{13} \epsilon^*_{24}}e^{\frac{2\pi}{\beta}\left( t-\frac{|\vec x|}{v_B} \right) }}\right)^{\Delta_V}
\label{eq-OTOC-BTZ}
\ee
where $\Delta_V$ and $\Delta_W$ are the scaling dimensions of the operator $V$ and $W$, respectively, and $\epsilon_{ij}=i(e^{i\epsilon_i}-e^{i \epsilon_j})$. For this system $\beta = 2\pi$ and $v_B=1$. This formula matches the direct CFT calculation\footnote{The CFT perspective for the onset of chaos has been widely discussed in \cite{Perlmutter:2016pkf}. Other references in this direction include, for instance \cite{Fitzpatrick:2016thx,Caputa:2016tgt,Turiaci:2016cvo,Caputa:2017rkm}.} obtained in \cite{Roberts:2014ifa}. It can also be derived using the geodesic approximation for two-sided correlators in a shock wave background \cite{BHchaos1,Roberts:2014ifa}.

Expanding the above result for small values of $G_\mt{N} \,e^{\frac{2\pi}{\beta}\left( t-\frac{|\vec x|}{v_B} \right) }$ we obtain 
\be
\text{OTO}(t)=1-8 \pi i G_\mt{N} \frac{\Delta_V \Delta_W}{\epsilon_{13} \epsilon^*_{24}} e^{\frac{2\pi}{\beta}\left( t-\frac{|\vec x|}{v_B} \right) }\,,
\ee
Since $h(t,\vec x) \sim  G_\mt{N}\,e^{\frac{2\pi}{\beta}\left( t-\frac{|\vec x|}{v_B} \right) }$, the above result implies 
\be
C(t,\vec x) \sim h(t,\vec x).
\ee 
The above result is valid for small\footnote{In AdS/CFT the Newton constant is related to the rank of the gauge group of dual CFT as $G_\mt{N} \sim 1/N^a$, where $a$ is a positive number that depends on the dimensionality of the bulk space time (cf. section 7.2 of \cite{Taylor:2017dly}). Our classical gravity calculations are only valid in the large-$N$ limit (that suppresses quantum corrections) so it is natural to consider $G_\mt{N}$ as a small parameter.} values of $G_\mt{N}$, or for any value of $G_\mt{N}$, but for times in the range $t_d <<t <<t_*$, where $t_* =\frac{\beta}{2\pi}\log \frac{1}{G_\mt{N}}$.

Despite being true in the Rindler AdS$_3$ case, the proportionality between the double commutator and the shock wave profile has not been demonstrated in more general cases. However, the authors of \cite{BHchaos4} argued that, in regions of moderate scattering between the V- and W-particle, the identification $C(t, \vec x) \sim h(t, \vec x)$ is approximately valid.

At very late times, the behavior of the OTO$(t)$ is expected to be controlled by the black hole quasi-normal modes. Indeed, in the case of a compact space, it is possible to show that 
\be
C(t) \sim e^{-2i \omega \left(t-t_*-R/v_B\right)}\,,\,\,\text{with}\,\,\text{Im}(\omega)<0\,,
\ee
where $R$ is the diameter of the compact space and $\omega$ is the system lowest quasi-normal frequency \cite{BHchaos4}.
\subsubsection{Stringy corrections}
In this section, we briefly discuss the effects of stringy corrections to the Einstein gravity results for OTOCs.
We start by reviewing the Einstein gravity results from the perspective of scattering amplitudes. In the framework of the Eikonal approximation, the phase shift suffered by the V-particle is given by
\be
\delta = - P^{V} h(t,\vec x) \sim G_\mt{N}\,s\,,
\ee
where we used the fact that $h(t,\vec x) \sim G_\mt{N}\,P^U$ and introduced a Mandelstam-like variable $s =2 A(0) P^U P^V$.
In a small-$G_\mt{N}$ expansion the double commutator $C(t)$ and the phase shift $\delta$ scale with $s$ in the same way, namely
\be
C(t) \sim G_\mt{N}s\,,
\ee
where $s\sim \beta^{-2}e^{\frac{2\pi}{\beta}t}$.

The string corrections can be incorporated using the standard Veneziano formula for the relativistic scattering amplitude $ \mathcal{A} \sim s\, \delta$. The phase shift can then be schematically written as an infinite sum 
\be
\delta \sim \sum_J G_\mt{N} s^{J-1}\,,
\ee
where each term correspond to the contribution due to the exchange of a spin-$J$ field. In Einstein gravity the dominant contribution comes from the exchange of a spin-2 field, the graviton. In string theory, we have to include an infinite tower of higher spin fields. Naively, it looks like these higher spin contributions will increase the development of chaos. However, the re-summation of the above sum actually leads to a decrease in the development of chaos. The string-corrected phase shift has a milder dependence with $s$, namely
\be
\delta \sim G_\mt{N}s^{J_\mt{eff}-1}\,,
\ee
with the effective spin given by \cite{BHchaos4}
\be
J_\mt{eff}=2-\frac{d(d-1)\ell_s^2}{4\ell_{AdS}^2}\,
\ee
where $\ell_s$ is the string length, $\ell_{AdS}$ is the AdS length scale and $d$ is the number of dimensions of the boundary theory. As a result, the string-corrected double commutator grows in time with an effective smaller Lyapunov exponent
\be
C_\mt{string}(t) \sim \,e^{\frac{2\pi}{\beta}\left(1-\frac{d(d-1)\ell_s^2}{4\ell_{AdS}^2} \right)t}\,,
\ee 
and this leads to a larger scrambling time\footnote{At small scales, the string-corrected shock wave has a gaussian profile, and the concept of butterfly velocity is not meaningful. It was recently shown, however, that at larger scales is possible to define a string-corrected butterfly velocity. The result for $\mathcal{N}=4$ SYM theory reads \cite{VB-string} $v_B =\sqrt{\frac{2}{3}}\left(1+\frac{23 \zeta(3)}{16}\frac{1}{\lambda^{3/2}} \right)$, where $\lambda$ is the 't Hooft coupling, which can be written in terms of string length scale as $\lambda=(\ell_{AdS}/\ell_s)^4$.}
\be
t^\mt{string}_*=t_*\left(1+\frac{d(d-1)\ell_s^2}{4\ell_{AdS}^2}\right)\,.
\ee

The above discussion implies that for a theory with a finite number of high-spin fields ($J>2$) chaos would develop faster than in Einstein gravity. These theories, however, are known to violate causality \cite{Camanho:2014apa}. It is then natural to speculate that the Lyapunov exponent obtained in Einstein gravity has the maximal possible value allowed by causality. This is indeed true and this is the topic of the next section.

\subsubsection{Bounds on chaos}

 One of the remarkable insights that came from the holographic description of quantum chaos is the fact that there is a bound on chaos - the quantum Lyapunov exponent is bounded from above, while the scrambling time is bounded from below. A distinct feature of holographic systems is that they saturate these two bounds.

Let us follow the historical order and start by discussing the lower bound on the scrambling time.
In black holes physics, the scrambling time defines how fast the information that has fallen into a black hole can be recovered from the emitted Hawking radiation\footnote{This assumes that half of the black hole's initial entropy has been radiated \cite{scrambling1}.}. In the context of the Hayden-Preskill thought experiment, the scrambling time is barely compatible with black hole complementarity \cite{scrambling1}, since a smaller scrambling time would lead to a violation of the non-clonning principle. This led Susskind and Sekino to conjecture that black holes are the fastest scramblers in nature, i.e., they have the smallest possible scrambling time \cite{scrambling2}. The lower bound on the scrambling time of a generic many-body quantum system can be written as
\begin{equation}
    t_* \geq C(\beta) \log N_\textrm{dof}
\end{equation}
where $C(\beta)$ is some function of the inverse temperature. In the case of black holes this function is simply given by $C(\beta)=\frac{\beta}{2\pi}$.

The scrambling time defines a stronger notion of thermalization, and should not be confused with the dissipation time. In fact, for black holes, one expects the dissipation time to be given by the black hole quasinormal modes\footnote{This is true in the case of low dimension operators.} $t_d \sim \beta$, while the scrambling time is parametrically larger $t_* \sim \beta \log N_\textrm{dof}$. This bring us to the second bound on chaos: for systems with such a large hierarchy between the scrambling and the dissipation time is possible to derive an upper bound for the Lyapunov exponent \cite{bound-chaos}
\begin{equation}
\lambda_L \leq \frac{2\pi}{\beta}\,.
\end{equation}
One should emphasize that this bound does not depend on the existence of a holographic dual. It can be derived for generic many-body quantum systems under some very reasonable assumptions.

The fact that black holes always have a maximum Lyapunov exponent led to the speculation that the saturation of the chaos bound might be a sufficient condition for a system to have an Einstein gravity dual \cite{bound-chaos,Kitaev-2014}. In fact, there have been many attempts to use the saturation of the chaos bound as a criterion to discriminate holographic CFTs from the non-holographic ones \cite{Roberts:2014ifa,Perlmutter:2016pkf,Fitzpatrick:2016thx,Caputa:2016tgt,Turiaci:2016cvo,Caputa:2017rkm,Michel:2016kwn,Padmanabhan:2017ekk}. It was recently shown, however, that this criterion, though necessary, is insufficient to guarantee a dual description purely in terms of Einstein gravity \cite{deBoer:2017xdk,Banerjee:2018twd}.

Since $v_B$ defines the speed at which information propagates it is natural to question whether this quantity is also bounded. From the perspective of the boundary theory, causality implies
\begin{equation}
    v_B \leq 1\,,
\end{equation}
meaning that information should not propagate faster than the speed of light. Indeed, the above bound can be derived in the context of Einstein gravity by using Null Energy Condition (NEC) and assuming an asymptotically AdS geometry\footnote{This derivation uses an alternative definition for $v_B$, which is based on entanglement wedge subregion duality \cite{Mezei-2016}.} \cite{Qi-2017}.  This is consistent with the expectation that gravity theories in asymptotically AdS geometries are dual to relativistic theories. In contrast, for geometries which are not asymptotically AdS, the butterfly velocity can surpass the speed of light \cite{Qi-2017,Fischler:2018kwt}, which is consistent with the non-Lorentz invariance of the corresponding boundary theories.

If we further assume isotropy, it is possible to derive a stronger bound for $v_B$ \cite{Mezei-2016v2}
\begin{equation}
    v_B \leq v_B^\textrm{Sch}=\sqrt{\frac{d}{2(d-1)}}\,,
    \label{eq-strongboundVB}
\end{equation}
where $v_B^\textrm{Sch}$ is the value of the butterfly velocity for an AdS-Schwarzschild black brane in $d+1$ dimensions. This is also the butterfly velocity for a $d$-dimensional thermal CFT.

The above formula shows that, for thermal CFTs, $v_B$ does not depend on the temperature. However, if we deform the CFT, $v_B$ acquires a temperature dependence as we move along the corresponding renormalization group (RG) flow. In fact, by considering deformations that break the rotational symmetry it was noticed that the butterfly velocity violates the above bound, but remains bounded from above by its value at the infra-red (IR) fixed point, never surpassing the speed of light \cite{Giataganas:2017koz,Jahnke-2017,Avila-2018}. The above bound can also be violated by higher curvature corrections, but $v_B$ remains bounded by the speed of light as long as causality is respected\footnote{For instance, in 4-dimensional Gauss-Bonnet (GB) gravity, the butterfly velocity surpasses the speed of light for $\lambda_{GB}<-3/4$, but causality requires $\lambda_{GB}>-0.19$ \cite{Camanho:2009vw,Buchel:2009sk}. Moreover, it was recently shown that, unless one adds an infinite tower of extra higher spin fields, GB gravity might be inconsistent with causality for any value of the GB coupling \cite{Camanho:2014apa}.}. The violation of the bound given in (\ref{eq-strongboundVB}) by anisotropy or higher curvature corrections is reminiscent of the well-known violation of the shear viscosity to entropy density ratio bound \cite{Brigante:2007nu,Brigante:2008gz,Camanho:2010ru,Erdmenger:2010xm,Rebhan:2011vd,Jahnke:2014vwa}.

\subsection{Chaos and Entanglement Spreading }

The thermofield double state displays a very atypical left-right pattern of entanglement that results from non-zero correlations between subsystems of QFT$_L$ and QFT$_R$ at $t=0$. The chaotic nature of the boundary theories is manifest by the fact that small perturbations added to the system in the asymptotic past destroy this delicate correlations \cite{BHchaos1}.

The special pattern of entanglement can be efficiently diagnosed by considering the mutual information $I(A,B)$ between spatial subsystems $A \subset$ QFT$_L$ and  $B \subset$ QFT$_R$, defined as
\begin{equation}
I(A,B)=S_A+S_B-S_{A \cup B}\,,
\end{equation}
where $S_A$ is the entanglement entropy of the subsystem $A$, and so on. The mutual information is always positive and provides an upper bound for correlations between operators $\mathcal{O}_L$ and $\mathcal{O}_R$ defined on $A$ and $B$, respectively \cite{boundIAB}
\begin{equation}
    I(A,B) \geq \frac{\left( \langle \mathcal{O}_L \mathcal{O}_R \rangle-\langle \mathcal{O}_L \rangle \langle \mathcal{O}_R \rangle \right)^2}{2 \langle \mathcal{O}_L^2 \rangle \langle \mathcal{O}_R^2 \rangle}\,.
\label{eq-Ibound}
\end{equation}

The thermofield double state has non-zero mutual information between large\footnote{For small subsystems, the mutual information is zero.} subsystems of the left and right boundary, signaling the existence of left-right correlations. These correlations can be destroyed by small perturbations in the asymptotic past, meaning that an initially positive mutual information drops to zero when we add a very early perturbation to the system.

Interestingly, the vanishing of the mutual information can be connected to the vanishing of the OTOCs discussed earlier.
If, for simplicity, we assume that $\mathcal{O}_L$ and $\mathcal{O}_R$ have zero thermal one point function, then the disruption of the mutual information implies the vanishing of the following four-point function
\begin{equation}
    \langle \mathcal{O}_L \mathcal{O}_R \rangle_W = \langle\textrm{TFD} |W_R^{\dagger} \mathcal{O}_L \mathcal{O}_R W_R|\textrm{TFD}\rangle =0\,,
\end{equation}
which is related by analytic continuation to the one-sided out-of-time-order correlator introduced earlier\footnote{To obtain an OTOC with operators acting only on the right boundary theory one just need to add $i \beta/2$ to time argument of the operator $\mathcal{O}_L$ in the above formula.}.

\begin{figure}[t!]
\begin{center}
\includegraphics[width=17cm]{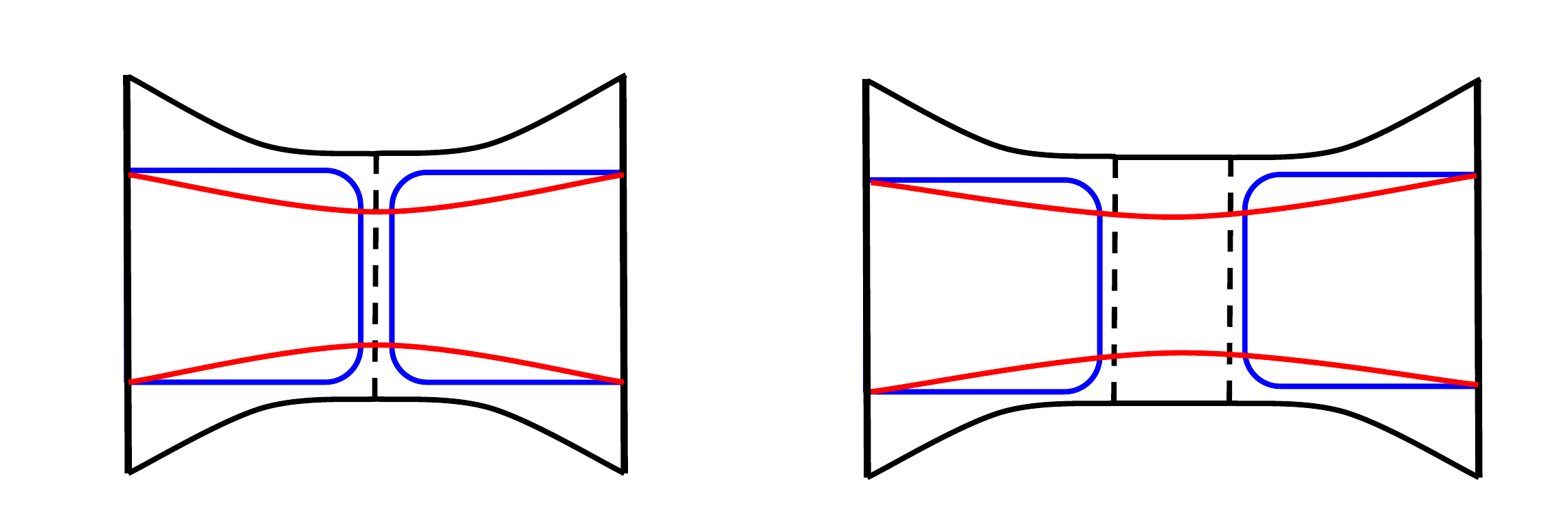}
         \put(-190,108){$\gamma_A$}
         \put(-60,110){$\gamma_B$}
         \put(-320,110){$\gamma_B$}
         \put(-420,110){$\gamma_A$}
         \put(-420,92){$\gamma_1$}
         \put(-420,52){$\gamma_2$}
         \put(-190,90){$\gamma_1$}
         \put(-190,50){$\gamma_2$}
         \put(-457,45){\small \rotatebox{90}{$\text{left boundary}$}}
         \put(-288,110){\small \rotatebox{-90}{$\text{right boundary}$}}
         \put(-230,45){\small \rotatebox{90}{$\text{left boundary}$}}
         \put(-25,110){\small \rotatebox{-90}{$\text{right boundary}$}}
         \put(-385,128){$\text{Horizon}$}
         \put(-370,125){\rotatebox{-90}{$\rightarrow$}}
         \put(-140,125){$\text{Horizons}$}
          \put(-140,123){\rotatebox{-135}{$\rightarrow$}}
          \put(-115,119){\rotatebox{-45}{$\rightarrow$}}       
         \put(-125,50){\small \rotatebox{90}{$\text{wormhole}$}}

\end{center}
\caption{ \small Illustration of the entangling surfaces in the $t=0$ slice of a two-sided black brane geometry. The U-shaped surfaces ($\gamma_A$ and $\gamma_B$) are represented by blue curves. The surface stretching through the wormhole is given by the union of the two red surfaces $\gamma_\textrm{wormhole}=\gamma_1 \cup \gamma_2$. In the {\it left panel} we represent the unperturbed geometry, in which the two horizons coincide. In the {\it right panel} we represent the geometry in the presence of a shock wave added at some time $t_0$ in the past. In this case, the size of the wormhole is effectively larger, and the two horizons no longer coincide.}
\label{fig-surfaces}
\end{figure}
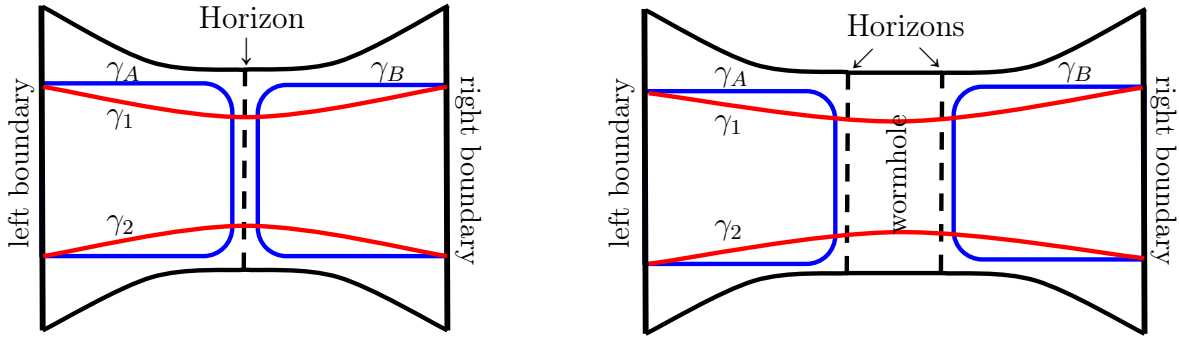

The disruption of the mutual information has a very simple geometrical realization in the bulk. The entanglement entropies that appear in the definition of $I(A,B)$ can be holographically calculated using the HRRT prescription \cite{RT,HRT}
\begin{equation}
    S_A =\frac{\textrm{Area}(\gamma_A)}{4 G_{N}}\,,
\end{equation}
where $\gamma_A$ is an extremal surface whose boundary coincides with the boundary of the region $A$. There is an analogous formula for $S_B$. Both $\gamma_A$ and $\gamma_B$ are U-shaped surfaces lying outside of the event horizon, in the left and right side of the geometry, respectively.
There are two candidates for the extremal surface that computes $S_{A \cup B}$: the surface $\gamma_A \cup \gamma_B$, or the surface $\gamma_\mt{wormhole}$ that connects the two asymptotic boundaries of the geometry. See figure \ref{fig-surfaces}. According to the RT prescription, we should pick the surface with less area. If $\gamma_A \cup \gamma_B$ has less area than $\gamma_\mt{wormhole}$, then $I(A,B)=0$, because Area($\gamma_A \cup \gamma_B$)=Area($\gamma_A$)+Area($\gamma_B$). On the other hand, if $\gamma_\mt{wormhole}$ has less area than $\gamma_A \cup \gamma_B$, i.e., Area($\gamma_\mt{wormhole}$) $<$ Area($\gamma_A$)+Area($\gamma_B$), then we have a positive mutual information 
\be
I(A,B)=\frac{1}{4\,G_\mt{N}} \left[ \text{Area}(\gamma_A)+\text{Area}(\gamma_B)-\text{Area}(\gamma_\mt{wormhole})\right] > 0\,.
\ee
Now, an early perturbation of the thermofield double state gives rise to a shock wave geometry in which the wormhole becomes longer. As a consequence the area of the surface $\gamma_\mt{wormhole}$ increases, resulting in a smaller mutual information. It is then clear that the mutual information will drop to zero if the wormhole is long enough. The length of the wormhole depends on the strength of the shock wave, which, by its turn, depends on how early is the perturbation producing it. Therefore, an early enough perturbation will produce a very long wormhole in which the mutual information will be zero.   The fact that the shock wave geometry produces a longer wormhole (along the $t=0$ slice of the geometry) is clearly seen if we represent the shock wave geometry with a tilted Penrose diagram. See, for instance, the figure 3 of \cite{Leichenauer-2014}.
%%%%%%%%%%%%%%%%%%%%%%%%%%%%%%%%%%%%%%%%%%%%%%%%%%%%
%%%%%%%%%%%%%%%%%%%%%%%%%%%%%%%%%%%%%%%%%%%%%%%%%%%%%

The mutual information $I(A,B)$ decreases as a function of the time $t_0$ at which we perturbed the system. For $t_0 \gtrsim t_*$, the mutual information decreases linearly with behavior controlled by the so-called entanglement velocity $v_E$ \cite{Jahnke-2017}
\be
\frac{d I(A,B)}{dt_0}=-\frac{d S_{A \cup B}}{dt_0}=-v_E \, s_\mt{th} \, \text{Area}(A \cup B)\,,
\ee
where $s_\mt{th}$ is the thermal entropy density and $\text{Area}(A \cup B)$ is the area of $A \cup B$ (or the volume of the boundary of this region). The two-sided black hole geometry with a shock wave can be thought of as an additional example of a holographic quench protocol \cite{Jahnke-2017}, and the time-dependence of entanglement entropy can be understood in terms of the so-called `entanglement tsunami' picture. See \cite{Calabrese:2005in} for field theory calculations, and \cite{AbajoArrastia:2010yt,Albash:2010mv,HM,tsunami1,tsunami2} for holographic calculations. However, it was recently shown that the entanglement tsunami picture is not very sharp. See \cite{Mezei-2016} for further details. In \cite{tsunami1,tsunami2}, the entanglement velocity was conjectured to be bounded as
\be
v_E \leq v_E^\mt{Sch}=\frac{\sqrt{d}(d-1)^{\frac{1}{2}-\frac{1}{d}}}{\left[2(d-1) \right]^{1-\frac{1}{d}}}\,,
\ee
where $v_E^\mt{Sch}$ is the entanglement velocity for a $(d+1)$-dimensional Schwarzschild black brane or, equivalently, the value of $v_E$ for a $d$-dimensional thermal CFT. This bound can be derived in the context of Einstein gravity assuming: an asymptotically AdS geometry, isotropy and NEC \cite{Mezei-2016v2}. Just like in the case of $v_B$, the entanglement velocity in thermal CFTs does not depend on the temperature. But $v_E$ acquires a temperature dependence if we deform the CFT and move along the corresponding RG flow \cite{Jahnke-2017, Avila-2018}. In these cases, $v_E$ violate the above bound, but it remains bounded by its corresponding value at the IR fixed point, never surpassing the speed of light. 

One can also prove that the entanglement velocity is also bounded by the speed of light\footnote{See \cite{Kundu:2016cgh,Lokhande:2017jik} for a discussion about small subsystems.}. This can be done by using the positivity of the mutual information \cite{Casini:2015zua}, or using inequalities involving the relative entropy \cite{Hartman-2015}. More generally, the authors of \cite{Mezei-2016} conjecture that $v_E \leq v_B$, which implies the bound $v_E<1$ in the cases where $v_B$ is bounded. However, both \cite{Casini:2015zua,Hartman-2015} assumed that the theory is Lorentz invariant. In the case of non-Lorentz invariant theories (e.g. non-commutative gauge theories)  the entanglement velocity can surpass the speed of light. This has been verified both in holography calculations \cite{Fischler:2018kwt} and in field theory calculations \cite{Sabella-Garnier:2017svs}.

Finally, we mention that other concepts from information theory can also be used to diagnose chaos in holography. It has been shown, for instance, that the relative entropy is also a useful tool to diagnose chaotic behavior \cite{Nakagawa:2018kvo}. For a connection between chaos and computational complexity, see, for instance \cite{Magan:2018nmu,HosseiniMansoori:2017tsm}.

\subsection{Chaos $\&$ Hydrodynamics}
Recently, there has been a growing interest in the connection between chaos and hydrodynamics \cite{Blake1,Blake2,Davison:2016ngz,Blake:2017qgd,Grozdanov:2017ajz,Blake-2017,Grozdanov:2018atb,Blake:2018leo,Haehl:2018izb}. Here we briefly review some interesting connection between chaos and diffusion phenomena.

A longstanding goal of quantum condensed matter physics is to have a deeper understanding of the so-called `strange metals'. These are strongly correlated materials that do not have a description in terms of quasiparticles excitations and whose transport properties display a remarkable degree of universality. In \cite{Sachdev-97,Sachdev-99}  Sachdev and Damle proposed that such a universal behavior could be explained by a fundamental dissipative timescale
\be
\tau_P \sim \frac{\hbar}{2\pi k_B T}\,,
\ee
that would govern the transport in such systems.

Interestingly, the Lyapunov exponent defines a time scale $\tau_L =1/\lambda_L$, and the upper bound on $\lambda_L$ translates into a lower bound for $\tau_L$ that precisely coincides with $\tau_P$
\be
\tau_L \geq \frac{\hbar \beta}{2\pi k_B}\,,
\label{eq-boundtau}
\ee 
where we reintroduced $\hbar$ and the Boltzmann constant in the expression for the bound on the Lyapunov exponent\footnote{In systems of units where $\hbar$ and $k_B$ are not equal to one, the bound on the Lyapunov exponent reads $\lambda_L \leq \frac{2\pi k_B}{\hbar \beta}$.}. Holographic systems saturate the above bound, and this explains the universality observed in the transport properties of these systems.

A prototypical example of universality is the linear resistivity of strange metals. In \cite{Hartnoll-2014},  Hartnoll proposed that the linear resistivity could be explained by the existence of a universal lower bound on the diffusion constants related to the collective diffusion of charge and energy
\be
D \gtrsim \hbar v^2 /(k_B T)\,,
\label{eq-Hartnollbound}
\ee
where $v$ is some characteristic velocity of the system. As $D$ is inversely proportional to the resistivity, systems saturating the above bound would  display linear resistivity behavior\footnote{See \cite{Jeong-2018} for a recent successful holographic description of linear resistivity at high-temperature.}. 

One should think of (\ref{eq-Hartnollbound}) as a reformulation of the Kovtun-Son-Starinets (KSS) bound \cite{Kovtun-2004}
\be
\frac{\eta}{s} \geq \frac{1}{4\pi}\frac{\hbar}{k_B}\,,
\ee
which also relies on the idea of a fundamental dissipative timescale $\tau_L \sim \hbar/(k_B T)$ controlling transport in strongly interacting systems. Naively, the observed violations of the KSS bound would seem to indicate the existence of systems in which the bound (\ref{eq-boundtau}) is violated. The bound (\ref{eq-Hartnollbound}) saves the idea of a fundamental dissipative timescale by introducing an additional parameter in the game, namely, the characteristic velocity $v$. The fact that $\eta/s$ can be made arbitrarily small in some systems corresponds to the fact that the characteristic velocity is highly suppressed in those cases. See\cite{Blake1} for further details.

In \cite{Blake1,Blake2}  Blake proposed that, at least for holographic systems with particle-hole symmetry, the characteristic velocity $v$ should be replaced by the butterfly velocity. More precisely
\be
D_{c}\geq C_{c}v_{B}^{2}\tau_{L},
\label{bound1}
\ee
where $D_c$ is the electric diffusity and $C_{c}$ is a constant that depends on the universality class of theory. This proposal was motivated by the fact that both $D_c$ and $v_B$ are determined by the dynamics close to the black hole horizon in the aforementioned systems. Despite working well for systems where energy and charge diffuse independently, this proposal was shown to fail in more general cases \cite{Lucas-2016,Davison-2016,Baggioli-2016,Kim-2017,Mokhtari:2017vyz}. This is related to the fact that, in more general cases, the diffusion of energy and charge is coupled, and the corresponding transport coefficients are not given only in terms of the geometry close to the black hole horizon. Hence, there is no reason for these coefficients to be related to the butterfly velocity, which is always determined solely by the near-horizon geometry.

There is, however, a universal piece of the diffusivity matrix that can be related to the chaos parameters at infra-red fixed points. This is the thermal diffusion constant  \cite{Blake3}
\be
D_{T}\geq C_{T}v_{B}^{2}\tau_{L}\,,
\ee 
where $C_{T}$ is a universality constant (different from $C_{c}$). This proposal was shown to be valid even for systems with spatial anisotropy  \cite{Ahn-2017}.  The above relation is not well defined when the system's dynamical critical exponent $z$ is equal to one, but it can be extended this case\footnote{We thank Hyun-Sik Jeong for calling our attention to this.} \cite{Davison-2018}.

Finally, we mention that there is an interesting relation between chaos and hydrodynamics that manifest itself in the so-called `pole-skipping' phenomenon. See \cite{Grozdanov:2017ajz,Blake-2017,Grozdanov:2018atb} for further details.

\section{Closing remarks}

The holographic description of quantum chaos not only has provided new insights into the inner-workings of gauge-gravity duality, but it has also given insights outside the scope of holography: some examples include the characterization of chaos with OTOCs, the definition of a quantum Lyapunov exponent and the existence of a bound for chaos. 

The success of this new approach to quantum chaos explains the growing experimental interest that OTOCs have been received. Indeed, several protocols for measuring OTOCs have been proposed, and there are already a few experimental results. See \cite{OTOC-swingle} and references therein. 

Finally, one of the remarkable features of quantum chaos is level statistics described by random matrices. The fact that this is present in the infrared limit of the SYK model \cite{Cotler:2016fpe, Garcia-Garcia:2016mno, Garcia-Garcia:2017pzl} suggests that it should also be present in quantum black holes\footnote{We thank A.~M.~Garc\'ia-Garc\'ia for calling our attention to this.}, although this has not yet been verified \cite{Saad:2018bqo}.

\section*{Acknowledgments}

It is a pleasure to thank A.~M.~Garc\'ia-Garc\'ia, S.~Nicolis, S.~A.~H.~Mansoori, and N. Garcia-Mata for useful correspondence. I also would like to thank Hyun-Sik Jeong for useful discussions.
This work was supported in part by Basic Science Research Program through the National Research Foundation of Korea(NRF) funded by the Ministry of Science, ICT \& Future Planning(NRF2017R1A2B4004810) and GIST Research Institute(GRI) grant funded by the GIST in 2018.

\section*{Conflicts of Interest}
The author declares that there is no conflict of interest regarding the publication of this paper.

%%%%%% BIBLIOGRAPHY


\begin{thebibliography}{99}

%%1
\bibitem{ullmo-2014}
D.~Ullmo, S.~Tomsovic,
``Introduction to quantum chaos'', http://www.lptms.u-psud.fr/membres/ullmo/Articles/eolss-ullmo-tomsovic.pdf, (2014)


%%2
\bibitem{duality1}
J.~M.~Maldacena,
``The large $N$ limit of superconformal field theories and supergravity,''
Adv.\ Theor.\ Math.\ Phys.\ {\bf 2}, 231 (1998)
[Int.\ J.\ Theor.\ Phys.\ {\bf 38}, 1113 (1999)]
[hep-th/9711200].

%%3
\bibitem{duality2}
  S.~S.~Gubser, I.~R.~Klebanov, A.~M.~Polyakov,
  ``Gauge theory correlators from noncritical string theory,''
  Phys.\ Lett.\  {\bf B428}, 105-114 (1998) [hep-th/9802109].

%%4
\bibitem{duality3}
  E.~Witten,
  ``Anti-de Sitter space and holography,''
  Adv.\ Theor.\ Math.\ Phys.\  {\bf 2}, 253-291 (1998) [hep-th/9802150].

%%5
 \bibitem{BHchaos1}
  S.~H.~Shenker and D.~Stanford,
  ``Black holes and the butterfly effect,''
  JHEP {\bf 1403}, 067 (2014)
  %doi:10.1007/JHEP03(2014)067
  [arXiv:1306.0622 [hep-th]].

%%6
\bibitem{BHchaos2}
  S.~H.~Shenker and D.~Stanford,
  ``Multiple Shocks,''
  JHEP {\bf 1412}, 046 (2014)
  %doi:10.1007/JHEP12(2014)046
  [arXiv:1312.3296 [hep-th]].

%%7
\bibitem{BHchaos3}
  D.~A.~Roberts, D.~Stanford and L.~Susskind,
  ``Localized shocks,''
  JHEP {\bf 1503}, 051 (2015)
  %doi:10.1007/JHEP03(2015)051
  [arXiv:1409.8180 [hep-th]].

%%8
\bibitem{BHchaos4}
  S.~H.~Shenker and D.~Stanford,
  ``Stringy effects in scrambling,''
  JHEP {\bf 1505}, 132 (2015)
  %doi:10.1007/JHEP05(2015)132
  [arXiv:1412.6087 [hep-th]].

%%9
\bibitem{Kitaev-2014}
  A.~Kitaev,
  ``Hidden Correlations in the Hawking Radiation and Thermal Noise,''
  talk given at Fundamental Physics Prize Symposium, Nov. 10, 2014.
Stanford SITP seminars, Nov. 11 and Dec. 18, 2014.

%%10
\bibitem{Rozenbaum-2017}
E.~B.~Rozenbaum, S.~Ganeshan and V.~Galitski,
``Lyapunov Exponent and Out-of-Time-Ordered Correlator's Growth Rate in a Chaotic System'',
Phys. Rev. Lett. {\bf 188}, 086801, 2017.

%%11
\bibitem{Rozenbaum-2018}
E.~B.~Rozenbaum, S.~Ganeshan and V.~Galitski,
``Universal Level Statistics of the Out-of-Time-Ordered Operator'',
arXiv:1801.10591 [cond-mat.dis-nn].

%%12
%\cite{Chavez-Carlos:2018ijc}
\bibitem{Chavez-Carlos:2018} 
  J.~Ch\'avez-Carlos, B.~L\'opez-Del-Carpio, M.~A.~Bastarrachea-Magnani, P.~Str\'ansk\'y, S.~Lerma-Hern\'andez, L.~F.~Santos and J.~G.~Hirsch,
  ``Quantum and Classical Lyapunov Exponents in Atom-Field Interaction Systems,''
  arXiv:1807.10292 [cond-mat.stat-mech].
  %%CITATION = ARXIV:1807.10292;%%
  %4 citations counted in INSPIRE as of 16 Nov 2018

%%%%%%%%%%%
%%%%%  AdS2-SYK
%\cite{Polchinski:2016xgd}
\bibitem{Polchinski:2016xgd} 
  J.~Polchinski and V.~Rosenhaus,
  ``The Spectrum in the Sachdev-Ye-Kitaev Model,''
  JHEP {\bf 1604}, 001 (2016)
  doi:10.1007/JHEP04(2016)001
  [arXiv:1601.06768 [hep-th]].
  %%CITATION = doi:10.1007/JHEP04(2016)001;%%
  %242 citations counted in INSPIRE as of 27 Nov 2018

%\cite{Maldacena:2016hyu}
\bibitem{Maldacena:2016hyu} 
  J.~Maldacena and D.~Stanford,
  ``Remarks on the Sachdev-Ye-Kitaev model,''
  Phys.\ Rev.\ D {\bf 94}, no. 10, 106002 (2016)
  doi:10.1103/PhysRevD.94.106002
  [arXiv:1604.07818 [hep-th]].
  %%CITATION = doi:10.1103/PhysRevD.94.106002;%%
  %403 citations counted in INSPIRE as of 27 Nov 2018

%\cite{Maldacena:2016upp}
\bibitem{Maldacena:2016upp} 
  J.~Maldacena, D.~Stanford and Z.~Yang,
  ``Conformal symmetry and its breaking in two dimensional Nearly Anti-de-Sitter space,''
  PTEP {\bf 2016}, no. 12, 12C104 (2016)
  doi:10.1093/ptep/ptw124
  [arXiv:1606.01857 [hep-th]].
  %%CITATION = doi:10.1093/ptep/ptw124;%%
  %215 citations counted in INSPIRE as of 27 Nov 2018
  
  %\cite{Kitaev:2017awl}
\bibitem{Kitaev:2017awl} 
  A.~Kitaev and S.~J.~Suh,
  ``The soft mode in the Sachdev-Ye-Kitaev model and its gravity dual,''
  JHEP {\bf 1805}, 183 (2018)
  doi:10.1007/JHEP05(2018)183
  [arXiv:1711.08467 [hep-th]].
  %%CITATION = doi:10.1007/JHEP05(2018)183;%%
  %73 citations counted in INSPIRE as of 27 Nov 2018



%%%%%%%%%%%%%%%%%%%%%%%%%%%%%%%%%%%%%%%%%%%%%%%%

%%% Stam Nicolis references %%%%%%%%%

%\cite{Axenides:2013iwa}
\bibitem{Axenides:2013iwa} 
  M.~Axenides, E.~G.~Floratos and S.~Nicolis,
  ``Modular discretization of the AdS$_{2}$/CFT$_{1}$ holography,''
  JHEP {\bf 1402}, 109 (2014)
  doi:10.1007/JHEP02(2014)109
  [arXiv:1306.5670 [hep-th]].
  %%CITATION = doi:10.1007/JHEP02(2014)109;%%
  %9 citations counted in INSPIRE as of 27 Nov 2018
  
  %\cite{Axenides:2015aha}
\bibitem{Axenides:2015aha} 
  M.~Axenides, E.~Floratos and S.~Nicolis,
  ``Chaotic Information Processing by Extremal Black Holes,''
  Int.\ J.\ Mod.\ Phys.\ D {\bf 24}, no. 09, 1542012 (2015)
  doi:10.1142/S0218271815420122
  [arXiv:1504.00483 [hep-th]].
  %%CITATION = doi:10.1142/S0218271815420122;%%
  %1 citations counted in INSPIRE as of 27 Nov 2018
  
  %\cite{Axenides:2016nmf}
\bibitem{Axenides:2016nmf} 
  M.~Axenides, E.~Floratos and S.~Nicolis,
  ``The quantum cat map on the modular discretization of extremal black hole horizons,''
  Eur.\ Phys.\ J.\ C {\bf 78}, no. 5, 412 (2018)
  doi:10.1140/epjc/s10052-018-5850-9
  [arXiv:1608.07845 [hep-th]].
  %%CITATION = doi:10.1140/epjc/s10052-018-5850-9;%%
  %1 citations counted in INSPIRE as of 27 Nov 2018
  

%%%%%%%%%%%%%%%%%%%%%%%%%%%%%%%%%%%%%%%%%
\bibitem{Roberts:2014ifa}
  D.~A.~Roberts and D.~Stanford,
  ``Two-dimensional conformal field theory and the butterfly effect,''
  Phys.\ Rev.\ Lett.\  {\bf 115}, no. 13, 131603 (2015)
  %doi:10.1103/PhysRevLett.115.131603
  [arXiv:1412.5123 [hep-th]].

%\cite{Stanford:2015owe}
\bibitem{Stanford:2015owe} 
  D.~Stanford,
  ``Many-body chaos at weak coupling,''
  JHEP {\bf 1610}, 009 (2016)
  doi:10.1007/JHEP10(2016)009
  [arXiv:1512.07687 [hep-th]].
  %%CITATION = doi:10.1007/JHEP10(2016)009;%%
  %56 citations counted in INSPIRE as of 27 Nov 2018

%\cite{Plamadeala:2018vsr}
\bibitem{Plamadeala:2018vsr} 
  E.~Plamadeala and E.~Fradkin,
  ``Scrambling in the quantum Lifshitz model,''
  J.\ Stat.\ Mech.\  {\bf 1806}, no. 6, 063102 (2018).
  doi:10.1088/1742-5468/aac136
  %%CITATION = doi:10.1088/1742-5468/aac136;%%

%%% randon unitary models
%\cite{Nahum:2017yvy}
\bibitem{Nahum:2017yvy} 
  A.~Nahum, S.~Vijay and J.~Haah,
  ``Operator Spreading in Random Unitary Circuits,''
  Phys.\ Rev.\ X {\bf 8}, no. 2, 021014 (2018)
  doi:10.1103/PhysRevX.8.021014
  [arXiv:1705.08975 [cond-mat.str-el]].
  %%CITATION = doi:10.1103/PhysRevX.8.021014;%%
  %67 citations counted in INSPIRE as of 27 Nov 2018

%\cite{Khemani:2017nda}
\bibitem{Khemani:2017nda} 
  V.~Khemani, A.~Vishwanath and D.~A.~Huse,
  ``Operator spreading and the emergence of dissipation in unitary dynamics with conservation laws,''
  Phys.\ Rev.\ X {\bf 8}, no. 3, 031057 (2018)
  doi:10.1103/PhysRevX.8.031057
  [arXiv:1710.09835 [cond-mat.stat-mech]].
  %%CITATION = doi:10.1103/PhysRevX.8.031057;%%
  %41 citations counted in INSPIRE as of 27 Nov 2018

%\cite{Rakovszky:2017qit}
\bibitem{Rakovszky:2017qit} 
  T.~Rakovszky, F.~Pollmann and C.~W.~von Keyserlingk,
  ``Diffusive hydrodynamics of out-of-time-ordered correlators with charge conservation,''
  Phys.\ Rev.\ X {\bf 8}, no. 3, 031058 (2018)
  doi:10.1103/PhysRevX.8.031058
  [arXiv:1710.09827 [cond-mat.stat-mech]].
  %%CITATION = doi:10.1103/PhysRevX.8.031058;%%
  %44 citations counted in INSPIRE as of 27 Nov 2018

%%% spin chain

%\cite{Luitz:2017jrn}
\bibitem{Luitz:2017jrn} 
  D.~J.~Luitz and Y.~Bar Lev,
  ``Information propagation in isolated quantum systems,''
  Phys.\ Rev.\ B {\bf 96}, no. 2, 020406 (2017)
  doi:10.1103/PhysRevB.96.020406
  [arXiv:1702.03929 [cond-mat.dis-nn]].
  %%CITATION = doi:10.1103/PhysRevB.96.020406;%%
  %28 citations counted in INSPIRE as of 27 Nov 2018

%\cite{Bohrdt:2016vhv}
\bibitem{Bohrdt:2016vhv} 
  A.~Bohrdt, C.~B.~Mendl, M.~Endres and M.~Knap,
  ``Scrambling and thermalization in a diffusive quantum many-body system,''
  New J.\ Phys.\  {\bf 19}, no. 6, 063001 (2017)
  doi:10.1088/1367-2630/aa719b
  [arXiv:1612.02434 [cond-mat.quant-gas]].
  %%CITATION = doi:10.1088/1367-2630/aa719b;%%
  %57 citations counted in INSPIRE as of 27 Nov 2018

\bibitem{Heyl-2018}
  M.~Heyl, F.~Pollman, and B.~D\'ora,
  ``Detecting Equilibrium and Dynamical Quantum Phase Transitions in Ising Chains via Out-of-Time-Ordered Correlators,''
  Phys. Rev. Lett. {\bf 121}, 016801 (2018)
  doi:10.1103/PhysRevLett.121.016801

\bibitem{Lin-2018}
  C-J.~Lin, O.~I.~Motrunich,
  ``Out-of-time-ordered correlators in quantum Ising chain,''
  Phys. Rev. B {\bf 97}, 144304 (2018)
  doi:10.1103/PhysRevB.97.144304

%\cite{Xu:2018xfz}
\bibitem{Xu:2018xfz} 
  S.~Xu and B.~Swingle,
  ``Accessing scrambling using matrix product operators,''
  arXiv:1802.00801 [quant-ph].
  %%CITATION = ARXIV:1802.00801;%%
  %24 citations counted in INSPIRE as of 27 Nov 2018






\bibitem{livro-chaos}
M.~Cencini, F.~Cecconi and A.~Vulpiani,
  ``Chaos: From Simple Models to Complex Systems,''
 World Scientific: Singapore, 2009.

%%%%%%%%%%%%%
%%13
%\bibitem{livro-chaos}
%M.~Cencini, F.~Cecconi and A.~Vulpiani,
%``Chaos: From Simple Models to Complex Systems,''
%World Scientiﬁc: Singapore, 2009.
%%%%

%\cite{Polchinski:2015cea}
\bibitem{Polchinski:2015cea} 
  J.~Polchinski,
  ``Chaos in the black hole S-matrix,''
  arXiv:1505.08108 [hep-th].
  %%CITATION = ARXIV:1505.08108;%%
  %69 citations counted in INSPIRE as of 27 Nov 2018

%\cite{Garcia-Mata:2018slr}
\bibitem{Garcia-Mata-2018} 
  I.~Garc\'ia-Mata, M.~Saraceno, R.~A.~Jalabert, A.~J.~Roncaglia and D.~A.~Wisniacki,
  ``Chaos signatures in the short and long time behavior of the out-of-time ordered correlator,''
  Phys.\ Rev.\ Lett.\  {\bf 121}, no. 21, 210601 (2018)
  doi:10.1103/PhysRevLett.121.210601
  [arXiv:1806.04281 [quant-ph]].
  %%CITATION = doi:10.1103/PhysRevLett.121.210601;%%
  %5 citations counted in INSPIRE as of 15 Jan 2019




\bibitem{Douglas-PITP}
D.~Stanford,
``Many-body quantum chaos,''
Seminar at the school {\it IAS PiTP 2018: From Qubtis to Spacetime}.
https://video.ias.edu/PiTP/2018/0723-DouglasStanford





%%14
\bibitem{larkin}
  A.~I.~Larkin and Y.~N.~Ovchinnikov,
  ``Quasiclassical method in the theory of superconductivity,''
  JETP 28, 6 (1969), 1200-1205.

%\cite{Khemani:2018sdn}
\bibitem{Khemani:2018sdn} 
  V.~Khemani, D.~A.~Huse and A.~Nahum,
  ``Velocity-dependent Lyapunov exponents in many-body quantum, semiclassical, and classical chaos,''
  Phys.\ Rev.\ B {\bf 98}, no. 14, 144304 (2018)
  doi:10.1103/PhysRevB.98.144304
  [arXiv:1803.05902 [cond-mat.stat-mech]].
  %%CITATION = doi:10.1103/PhysRevB.98.144304;%%
  %19 citations counted in INSPIRE as of 27 Nov 2018




%%15
\bibitem{Roberts:2016wdl}
  D.~A.~Roberts and B.~Swingle,
  ``Lieb-Robinson Bound and the Butterfly Effect in Quantum Field Theories,''
  Phys.\ Rev.\ Lett.\  {\bf 117}, no. 9, 091602 (2016)
  %doi:10.1103/PhysRevLett.117.091602
  [arXiv:1603.09298 [hep-th]].

%%16
\bibitem{eternalBH}
  J.~M.~Maldacena,
  ``Eternal black holes in anti-de Sitter,''
  JHEP {\bf 0304}, 021 (2003)
  %doi:10.1088/1126-6708/2003/04/021
  [hep-th/0106112].

%% 17
%\cite{Fidkowski:2003nf}
\bibitem{Fidkowski:2003nf} 
  L.~Fidkowski, V.~Hubeny, M.~Kleban and S.~Shenker,
  ``The Black hole singularity in AdS / CFT,''
  JHEP {\bf 0402}, 014 (2004)
  doi:10.1088/1126-6708/2004/02/014
  [hep-th/0306170].
  %%CITATION = doi:10.1088/1126-6708/2004/02/014;%%
  %242 citations counted in INSPIRE as of 16 Nov 2018

%% 18

  \bibitem{Dray-85}
  T.~Dray and G.~'t Hooft,
  ``The Gravitational Shock Wave of a Massless Particle,''
  Nucl.\ Phys.\ B {\bf 253}, 173 (1985).
  %doi:10.1016/0550-3213(85)90525-5

%% 19
\bibitem{Sfetsos-94}
  K.~Sfetsos,
  ``On gravitational shock waves in curved space-times,''
  Nucl.\ Phys.\ B {\bf 436}, 721 (1995)
  %doi:10.1016/0550-3213(94)00573-W
  [hep-th/9408169].

%%20
%\cite{Fischler:2018kwt}
\bibitem{Fischler:2018kwt} 
  W.~Fischler, V.~Jahnke and J.~F.~Pedraza,
  ``Chaos and entanglement spreading in a non-commutative gauge theory,''
  JHEP {\bf 1811}, 072 (2018)
  doi:10.1007/JHEP11(2018)072
  [arXiv:1808.10050 [hep-th]].
  %%CITATION = doi:10.1007/JHEP11(2018)072;%%
  %2 citations counted in INSPIRE as of 27 Nov 2018 

  %\cite{Baggioli:2018afg}
\bibitem{Baggioli-2018} 
  M.~Baggioli, B.~Padhi, P.~W.~Phillips and C.~Setty,
  ``Conjecture on the Butterfly Velocity across a Quantum Phase Transition,''
  JHEP {\bf 1807}, 049 (2018)
  doi:10.1007/JHEP07(2018)049
  [arXiv:1805.01470 [hep-th]].
  %%CITATION = doi:10.1007/JHEP07(2018)049;%%
  %4 citations counted in INSPIRE as of 16 Nov 2018 
%%21

  
%%22 
\bibitem{bound-chaos}
  J.~Maldacena, S.~H.~Shenker and D.~Stanford,
  ``A bound on chaos,''
  JHEP {\bf 1608}, 106 (2016)
  %doi:10.1007/JHEP08(2016)106
  [arXiv:1503.01409 [hep-th]]. 
 
%%23
\bibitem{Perlmutter:2016pkf}
  E.~Perlmutter,
  ``Bounding the Space of Holographic CFTs with Chaos,''
  JHEP {\bf 1610}, 069 (2016)
  %doi:10.1007/JHEP10(2016)069
  [arXiv:1602.08272 [hep-th]].


%%24
\bibitem{Fitzpatrick:2016thx}
  A.~L.~Fitzpatrick and J.~Kaplan,
  ``A Quantum Correction To Chaos,''
  JHEP {\bf 1605}, 070 (2016)
  %doi:10.1007/JHEP05(2016)070
  [arXiv:1601.06164 [hep-th]].

%%25
\bibitem{Caputa:2016tgt}
  P.~Caputa, T.~Numasawa and A.~Veliz-Osorio,
  ``Out-of-time-ordered correlators and purity in rational conformal field theories,''
  PTEP {\bf 2016}, no. 11, 113B06 (2016)
  %doi:10.1093/ptep/ptw157
  [arXiv:1602.06542 [hep-th]].


%%26
\bibitem{Turiaci:2016cvo}
  G.~Turiaci and H.~Verlinde,
  ``On CFT and Quantum Chaos,''
  JHEP {\bf 1612}, 110 (2016)
  %doi:10.1007/JHEP12(2016)110
  [arXiv:1603.03020 [hep-th]].

%%27
\bibitem{Caputa:2017rkm}
  P.~Caputa, Y.~Kusuki, T.~Takayanagi and K.~Watanabe,
  ``Out-of-Time-Ordered Correlators in $(T^2)^n/\mathbb{Z}_n$,''
  Phys.\ Rev.\ D {\bf 96}, no. 4, 046020 (2017)
  %doi:10.1103/PhysRevD.96.046020
  [arXiv:1703.09939 [hep-th]].

%\cite{Taylor:2017dly}
\bibitem{Taylor:2017dly} 
  M.~Taylor,
  ``Generalized conformal structure, dilaton gravity and SYK,''
  JHEP {\bf 1801}, 010 (2018)
  doi:10.1007/JHEP01(2018)010
  [arXiv:1706.07812 [hep-th]].
  %%CITATION = doi:10.1007/JHEP01(2018)010;%%
  %18 citations counted in INSPIRE as of 27 Nov 2018

%%%%%%%%%%%%%%%%%%%%%%%%%%%%%%%%%%%%%%%%%%%%%%%%%%%%%%%%%%%%%%%
%%%%%%%%%%%%%%%%%%%%%%%%%%%%%%%%%%%%%%%%%%%%%%%%%%%%%%%%%%%%%%%
%%%%%%%%%%%%%%%%%%%%%%%%%%%%%%%%%%%%%%%%%%%%%%%%%%%%%%%%%%%%%%%
%\cite{}
\bibitem{VB-string} 
  S.~Grozdanov,
  ``On the connection between hydrodynamics and quantum chaos in holographic theories with stringy corrections,''
  arXiv:1811.09641 [hep-th].
  %%CITATION = ARXIV:1811.09641;%%

%%%%%%%%%%%%%%%%%%%%%%%%%%%%%%%%%%%%%%%%%%%%%%%%%%%%%%%%%%%%%%
%%%%%%%%%%%%%%%%%%%%%%%%%%%%%%%%%%%%%%%%%%%%%%%%%%%%%%%%%%%%%%
%%%%%%%%%%%%%%%%%%%%%%%%%%%%%%%%%%%%%%%%%%%%%%%%%%%%%%%%%%%%%%

%%44
\bibitem{scrambling1}
  P.~Hayden and J.~Preskill,
  ``Black holes as mirrors: Quantum information in random subsystems,''
  JHEP {\bf 0709}, 120 (2007)
  %doi:10.1088/1126-6708/2007/09/120
  [arXiv:0708.4025 [hep-th]].

%%45
\bibitem{scrambling2}
  Y.~Sekino and L.~Susskind,
  ``Fast Scramblers,''
  JHEP {\bf 0810}, 065 (2008)
  %doi:10.1088/1126-6708/2008/10/065
  [arXiv:0808.2096 [hep-th]].




\bibitem{Camanho:2014apa}
  X.~O.~Camanho, J.~D.~Edelstein, J.~Maldacena and A.~Zhiboedov,
  ``Causality Constraints on Corrections to the Graviton Three-Point Coupling,''
  JHEP {\bf 1602}, 020 (2016)
  %doi:10.1007/JHEP02(2016)020
  [arXiv:1407.5597 [hep-th]].
















%%28
\bibitem{Michel:2016kwn}
  B.~Michel, J.~Polchinski, V.~Rosenhaus and S.~J.~Suh,
  ``Four-point function in the IOP matrix model,''
  JHEP {\bf 1605}, 048 (2016)
  %doi:10.1007/JHEP05(2016)048
  [arXiv:1602.06422 [hep-th]].

%\cite{Padmanabhan:2017ekk}
\bibitem{Padmanabhan:2017ekk} 
  P.~Padmanabhan, S.~J.~Rey, D.~Teixeira and D.~Trancanelli,
  ``Supersymmetric many-body systems from partial symmetries — integrability, localization and scrambling,''
  JHEP {\bf 1705}, 136 (2017)
  doi:10.1007/JHEP05(2017)136
  [arXiv:1702.02091 [hep-th]].
  %%CITATION = doi:10.1007/JHEP05(2017)136;%%
  %5 citations counted in INSPIRE as of 27 Nov 2018



%%29
\bibitem{deBoer:2017xdk}
  J.~de Boer, E.~Llabr\'es, J.~F.~Pedraza and D.~Vegh,
  ``Chaotic strings in AdS/CFT,''
  Phys.\ Rev.\ Lett.\  {\bf 120}, no. 20, 201604 (2018)
  %doi:10.1103/PhysRevLett.120.201604
  [arXiv:1709.01052 [hep-th]].

%%30
%\cite{Banerjee:2018twd}
\bibitem{Banerjee:2018twd} 
  A.~Banerjee, A.~Kundu and R.~R.~Poojary,
  ``Strings, Branes, Schwarzian Action and Maximal Chaos,''
  arXiv:1809.02090 [hep-th].
  %%CITATION = ARXIV:1809.02090;%%
  %2 citations counted in INSPIRE as of 16 Nov 2018
 
%%40
\bibitem{Qi-2017}
  X.~L.~Qi and Z.~Yang,
  ``Butterfly velocity and bulk causal structure,''
  arXiv:1705.01728 [hep-th]. 
 
 
%%31 
\bibitem{Mezei-2016v2}
  M.~Mezei,
  ``On entanglement spreading from holography,''
  JHEP {\bf 1705}, 064 (2017)
  %doi:10.1007/JHEP05(2017)064
  [arXiv:1612.00082 [hep-th]]. 
 
 %%32
 \bibitem{Giataganas:2017koz}
  D.~Giataganas, U.~G\"ursoy and J.~F.~Pedraza,
  ``Strongly-coupled anisotropic gauge theories and holography,''
  arXiv:1708.05691 [hep-th].

%%33
\bibitem{Jahnke-2017}
  V.~Jahnke,
  ``Delocalizing entanglement of anisotropic black branes,''
  JHEP {\bf 1801}, 102 (2018)
  %doi:10.1007/JHEP01(2018)102
  [arXiv:1708.07243 [hep-th]].

%\cite{Avila:2018sqf}
\bibitem{Avila-2018} 
  D.~Avila, V.~Jahnke and L.~Pati\~no,
  ``Chaos, Diffusivity, and Spreading of Entanglement in Magnetic Branes, and the Strengthening of the Internal Interaction,''
  JHEP {\bf 1809}, 131 (2018)
  doi:10.1007/JHEP09(2018)131
  [arXiv:1805.05351 [hep-th]].
  %%CITATION = doi:10.1007/JHEP09(2018)131;%%
  %3 citations counted in INSPIRE as of 15 Nov 2018
 
 
  
  %%%%%%%%%%%%%%%%%%%%%%%%%%%%%%%%%%%%%%%%%%%%%%%%%%%%%%%%%%%%%


%%41

\bibitem{Camanho:2009vw}
  X.~O.~Camanho and J.~D.~Edelstein,
  ``Causality constraints in AdS/CFT from conformal collider physics and Gauss-Bonnet gravity,''
  JHEP {\bf 1004}, 007 (2010)
  %doi:10.1007/JHEP04(2010)007
  [arXiv:0911.3160 [hep-th]].
%%42
\bibitem{Buchel:2009sk}
  A.~Buchel, J.~Escobedo, R.~C.~Myers, M.~F.~Paulos, A.~Sinha and M.~Smolkin,
  ``Holographic GB gravity in arbitrary dimensions,''
  JHEP {\bf 1003}, 111 (2010)
  %doi:10.1007/JHEP03(2010)111
  [arXiv:0911.4257 [hep-th]].
%%43

%%34
\bibitem{Brigante:2007nu}
  M.~Brigante, H.~Liu, R.~C.~Myers, S.~Shenker and S.~Yaida,
  ``Viscosity Bound Violation in Higher Derivative Gravity,''
  Phys.\ Rev.\ D {\bf 77}, 126006 (2008)
  %doi:10.1103/PhysRevD.77.126006
  [arXiv:0712.0805 [hep-th]].

%%35
\bibitem{Brigante:2008gz}
  M.~Brigante, H.~Liu, R.~C.~Myers, S.~Shenker and S.~Yaida,
  ``The Viscosity Bound and Causality Violation,''
  Phys.\ Rev.\ Lett.\  {\bf 100}, 191601 (2008)
  %doi:10.1103/PhysRevLett.100.191601
  [arXiv:0802.3318 [hep-th]].

%%36
\bibitem{Camanho:2010ru}
  X.~O.~Camanho, J.~D.~Edelstein and M.~F.~Paulos,
  ``Lovelock theories, holography and the fate of the viscosity bound,''
  JHEP {\bf 1105}, 127 (2011)
  %doi:10.1007/JHEP05(2011)127
  [arXiv:1010.1682 [hep-th]].

%%37
\bibitem{Erdmenger:2010xm}
  J.~Erdmenger, P.~Kerner and H.~Zeller,
  ``Non-universal shear viscosity from Einstein gravity,''
  Phys.\ Lett.\ B {\bf 699}, 301 (2011)
  %doi:10.1016/j.physletb.2011.04.009
  [arXiv:1011.5912 [hep-th]].

%%38
\bibitem{Rebhan:2011vd}
  A.~Rebhan and D.~Steineder,
  ``Violation of the Holographic Viscosity Bound in a Strongly Coupled Anisotropic Plasma,''
  Phys.\ Rev.\ Lett.\  {\bf 108}, 021601 (2012)
  %doi:10.1103/PhysRevLett.108.021601
  [arXiv:1110.6825 [hep-th]].

%%39
\bibitem{Jahnke:2014vwa}
  V.~Jahnke, A.~S.~Misobuchi and D.~Trancanelli,
  ``Holographic renormalization and anisotropic black branes in higher curvature gravity,''
  JHEP {\bf 1501}, 122 (2015)
  %doi:10.1007/JHEP01(2015)122
  [arXiv:1411.5964 [hep-th]].



%%%%%%%%%%%%%%%%%%%%%%%%%%%%%%%%%%%%%%%%%%%%%%%%
 \bibitem{boundIAB}
  M.~M.~ Wolf, F.~Verstraete, M.~B.~Hastings, and J.~I.~ Cirac,
   ``Area laws in quantum systems: Mutual information and correlations'',
  Phys.\ Rev.\ Lett.\  {\bf 100}, 070502 (2008)
  %doi:10.1103/PhysRevLett.100.070502
  [arXiv:0704.3906 [quant-ph]].

\bibitem{RT}
  S.~Ryu and T.~Takayanagi,
  ``Holographic derivation of entanglement entropy from AdS/CFT,''
  Phys.\ Rev.\ Lett.\  {\bf 96}, 181602 (2006)
  %doi:10.1103/PhysRevLett.96.181602
  [hep-th/0603001].

\bibitem{HRT}
  V.~E.~Hubeny, M.~Rangamani and T.~Takayanagi,
  ``A Covariant holographic entanglement entropy proposal,''
  JHEP {\bf 0707}, 062 (2007)
  %doi:10.1088/1126-6708/2007/07/062
  [arXiv:0705.0016 [hep-th]].





\bibitem{Leichenauer-2014}
  S.~Leichenauer,
  ``Disrupting Entanglement of Black Holes,''
  Phys.\ Rev.\ D {\bf 90}, no. 4, 046009 (2014)
  %doi:10.1103/PhysRevD.90.046009
  [arXiv:1405.7365 [hep-th]].



\bibitem{Sircar-2016}
  N.~Sircar, J.~Sonnenschein and W.~Tangarife,
  ``Extending the scope of holographic mutual information and chaotic behavior,''
  JHEP {\bf 1605}, 091 (2016)
  %doi:10.1007/JHEP05(2016)091
  [arXiv:1602.07307 [hep-th]].

\bibitem{Ling-2016}
  Y.~Ling, P.~Liu and J.~P.~Wu,
  ``Holographic Butterfly Effect at Quantum Critical Points,''
  JHEP {\bf 1710}, 025 (2017)
  %doi:10.1007/JHEP10(2017)025
  [arXiv:1610.02669 [hep-th]].

\bibitem{Cai-2017}
  R.~G.~Cai, X.~X.~Zeng and H.~Q.~Zhang,
  ``Influence of inhomogeneities on holographic mutual information and butterfly effect,''
  JHEP {\bf 1707}, 082 (2017)
  %doi:10.1007/JHEP07(2017)082
  [arXiv:1704.03989 [hep-th]].

\bibitem{Qaemmaqami-2017}
  M.~M.~Qaemmaqami,
  ``Criticality in third order lovelock gravity and butterfly effect,''
  Eur.\ Phys.\ J.\ C {\bf 78}, no. 1, 47 (2018)
  %doi:10.1140/epjc/s10052-018-5541-6
  [arXiv:1705.05235 [hep-th]].

\bibitem{Wu-2017}
  S.~F.~Wu, B.~Wang, X.~H.~Ge and Y.~Tian,
  ``Holographic RG flow of thermoelectric transport with momentum dissipation,''
  Phys.\ Rev.\ D {\bf 97}, no. 6, 066029 (2018)
  %doi:10.1103/PhysRevD.97.066029
  [arXiv:1706.00718 [hep-th]].

\bibitem{Qaemmaqami-2017-2}
  M.~M.~Qaemmaqami,
  ``Butterfly effect in 3D gravity,''
  Phys.\ Rev.\ D {\bf 96}, no. 10, 106012 (2017)
  %doi:10.1103/PhysRevD.96.106012
  [arXiv:1707.00509 [hep-th]].

\bibitem{Ahn-2017}
  H.~S.~Jeong, Y.~Ahn, D.~Ahn, C.~Niu, W.~J.~Li and K.~Y.~Kim,
  ``Thermal diffusivity and butterfly velocity in anisotropic Q-Lattice models,''
  JHEP {\bf 1801}, 140 (2018)
  %doi:10.1007/JHEP01(2018)140
  [arXiv:1708.08822 [hep-th]].





\bibitem{Calabrese:2005in}
  P.~Calabrese and J.~L.~Cardy,
  ``Evolution of entanglement entropy in one-dimensional systems,''
  J.\ Stat.\ Mech.\  {\bf 0504}, P04010 (2005)
  %doi:10.1088/1742-5468/2005/04/P04010
  [cond-mat/0503393].

\bibitem{AbajoArrastia:2010yt}
  J.~Abajo-Arrastia, J.~Aparicio and E.~Lopez,
  ``Holographic Evolution of Entanglement Entropy,''
  JHEP {\bf 1011}, 149 (2010)
  %doi:10.1007/JHEP11(2010)149
  [arXiv:1006.4090 [hep-th]].

\bibitem{Albash:2010mv}
  T.~Albash and C.~V.~Johnson,
  ``Evolution of Holographic Entanglement Entropy after Thermal and Electromagnetic Quenches,''
  New J.\ Phys.\  {\bf 13}, 045017 (2011)
  %doi:10.1088/1367-2630/13/4/045017
  [arXiv:1008.3027 [hep-th]].

\bibitem{HM}
  T.~Hartman and J.~Maldacena,
  ``Time Evolution of Entanglement Entropy from Black Hole Interiors,''
  JHEP {\bf 1305}, 014 (2013)
  %doi:10.1007/JHEP05(2013)014
  [arXiv:1303.1080 [hep-th]].

\bibitem{tsunami1}
  H.~Liu and S.~J.~Suh,
  ``Entanglement Tsunami: Universal Scaling in Holographic Thermalization,''
  Phys.\ Rev.\ Lett.\  {\bf 112}, 011601 (2014)
  %doi:10.1103/PhysRevLett.112.011601
  [arXiv:1305.7244 [hep-th]].

\bibitem{tsunami2}
  H.~Liu and S.~J.~Suh,
  ``Entanglement growth during thermalization in holographic systems,''
  Phys.\ Rev.\ D {\bf 89}, no. 6, 066012 (2014)
  %doi:10.1103/PhysRevD.89.066012
  [arXiv:1311.1200 [hep-th]].




%%%%%%%%%%%%%%%%%%%%%%%%%%%%%


\bibitem{Mezei-2016}
  M.~Mezei and D.~Stanford,
  ``On entanglement spreading in chaotic systems,''
  JHEP {\bf 1705}, 065 (2017)
  %doi:10.1007/JHEP05(2017)065
  [arXiv:1608.05101 [hep-th]].

\bibitem{Casini:2015zua}
  H.~Casini, H.~Liu and M.~Mezei,
  ``Spread of entanglement and causality,''
  JHEP {\bf 1607}, 077 (2016)
  %doi:10.1007/JHEP07(2016)077
  [arXiv:1509.05044 [hep-th]].

\bibitem{Hartman-2015}
  T.~Hartman and N.~Afkhami-Jeddi,
  ``Speed Limits for Entanglement,''
  arXiv:1512.02695 [hep-th].

\bibitem{Sabella-Garnier:2017svs}
  P.~Sabella-Garnier,
  ``Time dependence of entanglement entropy on the fuzzy sphere,''
  JHEP {\bf 1708}, 121 (2017)
  %doi:10.1007/JHEP08(2017)121
  [arXiv:1705.01969 [hep-th]].

\bibitem{Kundu:2016cgh}
  S.~Kundu and J.~F.~Pedraza,
  ``Spread of entanglement for small subsystems in holographic CFTs,''
  Phys.\ Rev.\ D {\bf 95}, no. 8, 086008 (2017)
  %doi:10.1103/PhysRevD.95.086008
  [arXiv:1602.05934 [hep-th]].

\bibitem{Lokhande:2017jik}
  S.~F.~Lokhande, G.~W.~J.~Oling and J.~F.~Pedraza,
  ``Linear response of entanglement entropy from holography,''
  JHEP {\bf 1710}, 104 (2017)
  %doi:10.1007/JHEP10(2017)104
  [arXiv:1705.10324 [hep-th]].

%\cite{Nakagawa:2018kvo}
\bibitem{Nakagawa:2018kvo} 
  Y.~O.~Nakagawa, G.~Sárosi and T.~Ugajin,
  ``Chaos and relative entropy,''
  JHEP {\bf 1807}, 002 (2018)
  doi:10.1007/JHEP07(2018)002
  [arXiv:1805.01051 [hep-th]].
  %%CITATION = doi:10.1007/JHEP07(2018)002;%%

%\cite{Magan:2018nmu}
\bibitem{Magan:2018nmu} 
  J.~M.~Magán,
  ``Black holes, complexity and quantum chaos,''
  JHEP {\bf 1809}, 043 (2018)
  doi:10.1007/JHEP09(2018)043
  [arXiv:1805.05839 [hep-th]].
  %%CITATION = doi:10.1007/JHEP09(2018)043;%%
  %11 citations counted in INSPIRE as of 27 Nov 2018

%\cite{HosseiniMansoori:2017tsm}
\bibitem{HosseiniMansoori:2017tsm} 
  S.~A.~Hosseini Mansoori and M.~M.~Qaemmaqami,
  ``Complexity Growth, Butterfly Velocity and Black hole Thermodynamics,''
  arXiv:1711.09749 [hep-th].
  %%CITATION = ARXIV:1711.09749;%%
  %18 citations counted in INSPIRE as of 27 Nov 2018















\bibitem{Blake1}
  M.~Blake,
  ``Universal Charge Diffusion and the Butterfly Effect in Holographic Theories,''
  Phys.\ Rev.\ Lett.\  {\bf 117}, no. 9, 091601 (2016)
  %doi:10.1103/PhysRevLett.117.091601
  [arXiv:1603.08510 [hep-th]].

\bibitem{Blake2}
  M.~Blake,
  ``Universal Diffusion in Incoherent Black Holes,''
  Phys.\ Rev.\ D {\bf 94}, no. 8, 086014 (2016)
  %doi:10.1103/PhysRevD.94.086014
  [arXiv:1604.01754 [hep-th]].

\bibitem{Davison:2016ngz}
  R.~A.~Davison, W.~Fu, A.~Georges, Y.~Gu, K.~Jensen and S.~Sachdev,
  ``Thermoelectric transport in disordered metals without quasiparticles: The Sachdev-Ye-Kitaev models and holography,''
  Phys.\ Rev.\ B {\bf 95}, no. 15, 155131 (2017)
  %doi:10.1103/PhysRevB.95.155131
  [arXiv:1612.00849 [cond-mat.str-el]].

\bibitem{Blake:2017qgd}
  M.~Blake, R.~A.~Davison and S.~Sachdev,
  ``Thermal diffusivity and chaos in metals without quasiparticles,''
  Phys.\ Rev.\ D {\bf 96}, no. 10, 106008 (2017)
  %doi:10.1103/PhysRevD.96.106008
  [arXiv:1705.07896 [hep-th]].

\bibitem{Grozdanov:2017ajz}
  S.~Grozdanov, K.~Schalm and V.~Scopelliti,
  ``Black hole scrambling from hydrodynamics,''
  Phys.\ Rev.\ Lett.\  {\bf 120}, no. 23, 231601 (2018)
  %doi:10.1103/PhysRevLett.120.231601
  [arXiv:1710.00921 [hep-th]].

\bibitem{Blake-2017}
  M.~Blake, H.~Lee and H.~Liu,
  ``A quantum hydrodynamical description for scrambling and many-body chaos,''
  arXiv:1801.00010 [hep-th].

\bibitem{Grozdanov:2018atb}
  S.~Grozdanov, K.~Schalm and V.~Scopelliti,
  ``Kinetic theory for classical and quantum many-body chaos,''
  arXiv:1804.09182 [hep-th].

%\cite{Blake:2018leo}
\bibitem{Blake:2018leo} 
  M.~Blake, R.~A.~Davison, S.~Grozdanov and H.~Liu,
  ``Many-body chaos and energy dynamics in holography,''
  JHEP {\bf 1810}, 035 (2018)
  doi:10.1007/JHEP10(2018)035
  [arXiv:1809.01169 [hep-th]].
  %%CITATION = doi:10.1007/JHEP10(2018)035;%%
  %1 citations counted in INSPIRE as of 16 Nov 2018

%\cite{Haehl:2018izb}
\bibitem{Haehl:2018izb} 
  F.~M.~Haehl and M.~Rozali,
  ``Effective Field Theory for Chaotic CFTs,''
  JHEP {\bf 1810}, 118 (2018)
  doi:10.1007/JHEP10(2018)118
  [arXiv:1808.02898 [hep-th]].
  %%CITATION = doi:10.1007/JHEP10(2018)118;%%
  %6 citations counted in INSPIRE as of 16 Nov 2018







  

 
\bibitem{Sachdev-97} 
 S.~Sachdev and K. ~Damle
 ``Non-zero temperature transport near quantum critical point'' 
   Phys.\ Rev.\ B {bf 56}, 8714 (1997). 
   [arXiv:cond-mat/9705206]

%% 35
\bibitem{Sachdev-99} 
  S.~Sachdev,
  ``Quantum phase transitions,''
  Cambrigde University Press (1999)
  
%% 36
%\cite{Hartnoll:2014lpa}
\bibitem{Hartnoll-2014} 
  S.~A.~Hartnoll,
  ``Theory of universal incoherent metallic transport,''
  Nature Phys.\  {\bf 11}, 54 (2015)
  doi:10.1038/nphys3174
  [arXiv:1405.3651 [cond-mat.str-el]].
  %%CITATION = doi:10.1038/nphys3174;%%
  %107 citations counted in INSPIRE as of 12 Mar 2018

%\cite{Jeong:2018tua}
\bibitem{Jeong-2018} 
  H.~S.~Jeong, K.~Y.~Kim and C.~Niu,
  ``Linear-$T$ resistivity at high temperature,''
  JHEP {\bf 1810}, 191 (2018)
  doi:10.1007/JHEP10(2018)191
  [arXiv:1806.07739 [hep-th]].
  %%CITATION = doi:10.1007/JHEP10(2018)191;%%
  %1 citations counted in INSPIRE as of 27 Jan 2019


%\cite{Kovtun:2004de}
\bibitem{Kovtun-2004} 
  P.~Kovtun, D.~T.~Son and A.~O.~Starinets,
  ``Viscosity in strongly interacting quantum field theories from black hole physics,''
  Phys.\ Rev.\ Lett.\  {\bf 94}, 111601 (2005)
  doi:10.1103/PhysRevLett.94.111601
  [hep-th/0405231].
  %%CITATION = doi:10.1103/PhysRevLett.94.111601;%%
  %2107 citations counted in INSPIRE as of 27 Jan 2019




%% 37
%\cite{Lucas:2016yfl}
\bibitem{Lucas-2016} 
  A.~Lucas and J.~Steinberg,
  ``Charge diffusion and the butterfly effect in striped holographic matter,''
  JHEP {\bf 1610}, 143 (2016)
  doi:10.1007/JHEP10(2016)143
  [arXiv:1608.03286 [hep-th]].
  %%CITATION = doi:10.1007/JHEP10(2016)143;%%
  %29 citations counted in INSPIRE as of 12 Mar 2018

%% 38
%\cite{Davison:2016ngz}
\bibitem{Davison-2016} 
  R.~A.~Davison, W.~Fu, A.~Georges, Y.~Gu, K.~Jensen and S.~Sachdev,
  ``Thermoelectric transport in disordered metals without quasiparticles: The Sachdev-Ye-Kitaev models and holography,''
  Phys.\ Rev.\ B {\bf 95}, no. 15, 155131 (2017)
  doi:10.1103/PhysRevB.95.155131
  [arXiv:1612.00849 [cond-mat.str-el]].
  %%CITATION = doi:10.1103/PhysRevB.95.155131;%%
  %96 citations counted in INSPIRE as of 12 Mar 2018

%% 39
%\cite{Baggioli:2016pia}
\bibitem{Baggioli-2016} 
  M.~Baggioli, B.~Goutéraux, E.~Kiritsis and W.~J.~Li,
  ``Higher derivative corrections to incoherent metallic transport in holography,''
  JHEP {\bf 1703}, 170 (2017)
  doi:10.1007/JHEP03(2017)170
  [arXiv:1612.05500 [hep-th]].
  %%CITATION = doi:10.1007/JHEP03(2017)170;%%
  %23 citations counted in INSPIRE as of 12 Mar 2018

%% 40
%\cite{Kim:2017dgz}
\bibitem{Kim-2017} 
  K.~Y.~Kim and C.~Niu,
  ``Diffusion and Butterfly Velocity at Finite Density,''
  JHEP {\bf 1706}, 030 (2017)
  doi:10.1007/JHEP06(2017)030
  [arXiv:1704.00947 [hep-th]].
  %%CITATION = doi:10.1007/JHEP06(2017)030;%%
  %18 citations counted in INSPIRE as of 12 Mar 2018

%\cite{Mokhtari:2017vyz}
\bibitem{Mokhtari:2017vyz} 
  A.~Mokhtari, S.~A.~Hosseini Mansoori and K.~Bitaghsir Fadafan,
  ``Diffusivities bounds in the presence of Weyl corrections,''
  Phys.\ Lett.\ B {\bf 785}, 591 (2018)
  doi:10.1016/j.physletb.2018.09.020
  [arXiv:1710.03738 [hep-th]].
  %%CITATION = doi:10.1016/j.physletb.2018.09.020;%%
  %7 citations counted in INSPIRE as of 27 Nov 2018


%% 41
%\cite{Blake:2017qgd}
\bibitem{Blake3} 
  M.~Blake, R.~A.~Davison and S.~Sachdev,
  ``Thermal diffusivity and chaos in metals without quasiparticles,''
  Phys.\ Rev.\ D {\bf 96}, no. 10, 106008 (2017)
  doi:10.1103/PhysRevD.96.106008
  [arXiv:1705.07896 [hep-th]].
  %%CITATION = doi:10.1103/PhysRevD.96.106008;%%
  %26 citations counted in INSPIRE as of 09 Mar 2018

%% 42
%\cite{Ahn:2017kvc}
\bibitem{Ahn-2017} 
  D.~Ahn, Y.~Ahn, H.~S.~Jeong, K.~Y.~Kim, W.~J.~Li and C.~Niu,
  ``Thermal diffusivity and butterfly velocity in anisotropic Q-Lattice models,''
  arXiv:1708.08822 [hep-th].
  %%CITATION = ARXIV:1708.08822;%% 
  
%\cite{Davison:2018nxm}
\bibitem{Davison-2018} 
  R.~A.~Davison, S.~A.~Gentle and B.~Goutéraux,
  ``Impact of irrelevant deformations on thermodynamics and transport in holographic quantum critical states,''
  arXiv:1812.11060 [hep-th].
  %%CITATION = ARXIV:1812.11060;%%  
  
  

 
 
 \bibitem{OTOC-swingle}
 B.~Swingle,
 ``Unscrambling the physics of out-of-time-order correlators'',
 Nature Physics, {\bf 14}, 988–990 (2018)
 
 
%%% Garcia-Garcia: SYK energy spectrum behaves like random matrix, hence, black holes also must have a manifestation of this 
 %\cite{Garcia-Garcia:2016mno}
 
%\cite{Cotler:2016fpe}
\bibitem{Cotler:2016fpe} 
  J.~S.~Cotler {\it et al.},
  ``Black Holes and Random Matrices,''
  JHEP {\bf 1705}, 118 (2017)
  Erratum: [JHEP {\bf 1809}, 002 (2018)]
  doi:10.1007/JHEP09(2018)002, 10.1007/JHEP05(2017)118
  [arXiv:1611.04650 [hep-th]].
  %%CITATION = doi:10.1007/JHEP09(2018)002, 10.1007/JHEP05(2017)118;%%
  %135 citations counted in INSPIRE as of 27 Nov 2018 
 
 
 
\bibitem{Garcia-Garcia:2016mno} 
  A.~M.~Garc\'ia-Garc\'ia and J.~J.~M.~Verbaarschot,
  ``Spectral and thermodynamic properties of the Sachdev-Ye-Kitaev model,''
  Phys.\ Rev.\ D {\bf 94}, no. 12, 126010 (2016)
  doi:10.1103/PhysRevD.94.126010
  [arXiv:1610.03816 [hep-th]].
  %%CITATION = doi:10.1103/PhysRevD.94.126010;%%
  %98 citations counted in INSPIRE as of 27 Nov 2018

%\cite{Garcia-Garcia:2017pzl}
\bibitem{Garcia-Garcia:2017pzl} 
  A.~M.~Garc\'ia-Garc\'ia and J.~J.~M.~Verbaarschot,
  ``Analytical Spectral Density of the Sachdev-Ye-Kitaev Model at finite N,''
  Phys.\ Rev.\ D {\bf 96}, no. 6, 066012 (2017)
  doi:10.1103/PhysRevD.96.066012
  [arXiv:1701.06593 [hep-th]].
  %%CITATION = doi:10.1103/PhysRevD.96.066012;%%
  %66 citations counted in INSPIRE as of 27 Nov 2018  
  
 
%%%%%%%%%%%%%%%%%%%%%%%%%%%%%%%%%%%%% 

%\cite{Saad:2018bqo}
\bibitem{Saad:2018bqo} 
  P.~Saad, S.~H.~Shenker and D.~Stanford,
  ``A semiclassical ramp in SYK and in gravity,''
  arXiv:1806.06840 [hep-th].
  %%CITATION = ARXIV:1806.06840;%%
  %10 citations counted in INSPIRE as of 27 Nov 2018




\end{thebibliography}
\end{document}